\documentclass[11pt,a4paper]{article}
\usepackage{jheppub}
\hypersetup{colorlinks=true,linkcolor=blue,citecolor=magenta,filecolor=magenta,
  urlcolor=cyan}

\usepackage{multirow}
\usepackage{float}
\usepackage{slashed}
\usepackage{tikz-feynman}

\definecolor{darkmagenta}{rgb}{0.55, 0.0, 0.55}
\definecolor{blue}{rgb}{0.0, 0.5, 0.69}

\begin{document}
\hspace*{112mm}{\large \tt CTPU-PTC-21-19} \\
\title{Contact interactions and top-philic scalar dark matter}

\author[a]{Alan S. Cornell}
\emailAdd{acornell@uj.ac.za}

\author[b]{\!\!, Aldo Deandrea}
\emailAdd{deandrea@ipnl.in2p3.fr}

\author[c]{\!\!, Thomas Flacke}
\emailAdd{flacke@ibs.re.kr}

\author[d,e]{\!\!, Benjamin Fuks}
\emailAdd{fuks@lpthe.jussieu.fr}

\author[a,b]{\!and Lara~Mason}
\emailAdd{mason@ipnl.in2p3.fr}

\affiliation[a]{Department of Physics, University of Johannesburg, PO Box 524,
  Auckland Park 2006, South Africa}
\affiliation[b]{Universit\'e de Lyon, F-69622 Lyon, France: Universit\'e Lyon 1,
  Villeurbanne CNRS/IN2P3, UMR5822, Institut de Physique des 2 Infinis de Lyon}
\affiliation[c]{Center for Theoretical Physics of the Universe, Institute for Basic Science (IBS), Daejeon 34126,Korea}
\affiliation[d]{Laboratoire de Physique Th\'eorique et Hautes Energies (LPTHE),
  UMR 7589, Sorbonne Universit\'e et CNRS, 4 place Jussieu,
  75252 Paris Cedex 05, France}
\affiliation[e]{Institut Universitaire de France, 103 boulevard Saint-Michel,
  75005 Paris, France}

\date{\today}

\abstract{We investigate the phenomenology of a scalar top-philic dark matter candidate when adding a dimension-five contact interaction term, as motivated by possible underlying extensions of the Standard Model such as composite Higgs models. We show that the presence of contact interactions can have a major impact on the dark matter relic density as well as on its direct and indirect detection prospects, while the collider phenomenology of the model is unaffected. This underlines the complementarity of collider and cosmological constraints on dark matter models.}

\maketitle
\flushbottom



\section{Introduction}

That about 27\% of the universe's energy budget is made up of matter which cannot be described by the Standard Model (SM) is one of the foremost mysteries in particle physics. Dark matter (DM), so named for being non-luminous and non-absorbing~\cite{Adam:2015rua}, is non-relativistic matter which does not behave like the baryonic matter of the SM. The nature and origin of DM is still unknown, despite compelling evidence~\cite{Bertone:2010zza} for its existence, and numerous physics programs have been established in an attempt to detect it. These include direct and indirect detection attempts and efforts at colliders such as the Large Hadron Collider (LHC) at CERN, none of which have yet made conclusive discoveries.

The ``cold'' DM theory postulates that DM has been (and continues to be) non-relativistic since the beginnings of galaxy formation, with its evolution governed by the Boltzmann equation. In the early universe, DM was in thermal equilibrium with SM particles. As the universe expanded and thereby cooled, the DM collision rate dropped to the Hubble expansion rate, at which point the particles ``froze out'' and decoupled. The observed DM density, or relic density $\Omega_{DM}h^2$, became constant. Candidates for cold non-baryonic DM are therefore strongly constrained by the relic density, which has been measured to be~\cite{Aghanim:2018eyx} 
\begin{equation}
\Omega_{DM}h^2 = 0.1186 \pm 0.0020,
\label{eq:relic}
\end{equation}
and which is controlled by the annihilation cross section of the DM candidate. A number of candidates for DM have been proposed, but the WIMP paradigm, in which a new dark state couples to the SM through a generic weak interaction and has a mass ranging from several GeV to a few TeV~\cite{Bertone:2010zza,Arcadi:2017kky}, persists as a promising avenue.

In this work we build on existing models of a scalar heavy DM candidate $S$ that manifests as a weakly interacting massive particle (WIMP) coupling to the SM top quark $t$ via a Yukawa-type term which involves a heavy fermionic mediator $T$. Moreover, we constrain the mass $m_S$ of our real scalar DM particle $S$ to be larger than that of the top quark ($m_t$), focusing on candidates $S$ with $200~{\rm GeV}~\lesssim~m_S~\lesssim 3~{\rm TeV}$. As we are keeping in mind that $S$ and $T$ could be resonances emerging from a composite Higgs theory, we require their masses to lie within an order of magnitude of each other. In this $m_S$ range the DM annihilation channel $SS\rightarrow tt$ dominates, and with $m_S$ well separated from the $m_t$ threshold effects which would arise in more mass-degenerate regimes can be avoided.

The DM resonance here considered is thus top-philic, as suggested in composite extensions of the SM in which the top quark plays a special role, and couples to the SM via a fermionic mediator $T$ which is also heavy. This occurs via a $t$-channel interaction originating from an $STt$ operator. Many scalar DM models including a $t$-channel fermionic mediator have previously been studied~\cite{ An:2013xka, Baek:2016lnv, Baek:2017ykw,Arina:2018zcq,Colucci:2018vxz,Colucci:2018qml,Arina:2020udz}, as top-philic DM models~\cite{Zhang:2012da, Batell:2013zwa, Kumar:2013hfa, Gomez:2014lva, Kilic:2015vka, Arina:2016cqj, Cheung:2010zf} and heavy DM with masses ranging up to several TeV have been discussed in the literature~\cite{Beneke:2018ssm}. In particular, recent investigations~\cite{Colucci:2018vxz,Colucci:2018qml} have shown that while leading order (LO) calculations allow for a heavy DM, next-to-leading-order (NLO) corrections to the annihilation cross section lead to significant modifications of the existing constraints on the model parameter space.

In the current study, we envision that the DM candidate $S$ and the fermionic mediator $T$ may arise within a composite Higgs model with an underlying fermionic construction as composite bound states. Such a composite bound state would be expected to have a mass of the order of the energy scale of the theory $\Lambda$, which could be expected to be several ${\rm TeV}$~\cite{Bellazzini:2012tv}. While the possibility that DM arises from a composite Higgs model as an additional pseudo-Nambu-Goldstone boson (pNGB) has been well studied, the prospect that it is a heavy resonance has received less attention. It is this structure which we bear in mind throughout this work, although the couplings and parameters are left free to allow for the discussion of a more general situation. With this aim of generality, we recall that dimension-five operators are a generic feature of a broad range of Beyond the Standard Model (BSM) theories. In particular, these include composite Higgs models arising from strong dynamics, where higher dimensional operators do not decouple~\cite{Bellazzini:2012tv} and may therefore be relevant at colliders and in direct and indirect detection experiments. Instead of focusing on a particular theory, our methodology complements and generalises earlier studies relying a simplified model construction~\cite{Colucci:2018vxz,Colucci:2018qml} and add to such a modelling an additional dimension-five operator $SStt$ which emerges from a contact term between the $S$ and $t$ states. This independent dimension-five operator comes with an unknown $\mathcal{O}(1)$ Wilson coefficient, whose sign and magnitude could substantially modify the existing limits obtained through only dimension-four operators.\

In fact, the addition of this term contributes to the relic density calculations by opening an area of parameter space not previously allowed, where the Yukawa term $\tilde{y}_t$ governing the $STt$ operator is very small. The interplay between the dimension-five operator and Yukawa term is fully parametrisable according to the masses of the particles, but is subject to possibly large interference for $m_S \approx m_T$. We begin this work by investigating the predictability of the behaviour of the system across mass compression scales and with the inclusion of the additional dimension-five term. We have determined the analytical function which predicts the interplay between the dimension-five operator and the Yukawa coefficient which results in the correct relic density, potentially bypassing the need for intensive numerical calculations in order to determine the parameters yielding the correct relic density. In doing so we also display the need to account for co-annihilations in the highly compressed regime.\

We then calculate expected bounds from DM direct detection experiments from the DM candidate $S$ scattering off atomic nuclei. While the absence of a valence top quark from the nucleus makes the DM-gluon interactions the only avenue of detection, we may expect some suppression due to the loop-generated nature of these interactions. We see in fact that the addition of the new interaction term greatly improves the direct detection prospects, where many potential model configurations live above the neutrino floor and are therefore hypothetically accessible. Additionally, some models are within reach of the XENON-1T experiment. In these discussions we consider both possible signs of the Wilson coefficient. In studying the indirect detection constraints and prospects, we also find that more of the parameter space becomes accessible. Finally, we examine the collider constraints relevant to our model, considering the reinterpretation of existing mono-jet and multi-jet analyses to assess our model against current bounds. We find no improvement on existing collider constraints due to the dimension-five effects, and instead update the bounds given in previous analyses in the light of full LHC run~2 results. We moreover estimate the sensitivity of the future high-luminosity phase of the LHC to the considered class of top-philic scalar DM models.

In the following we begin with a description of the theory in section~\ref{sec:theory}, defining the model and cross sections relevant to the relic density calculation. In order to map out the parameter space of interest, the relic density is calculated and relevant values of the Yukawa and dimension-five couplings are determined. In section~\ref{sec:directdetec} we study the impact on the direct detection constraints by the addition of this dimension-five term, while indirect detection implications are discussed in section \ref{sec:inddec}. In section~\ref{sec:collider} we describe the hadron collider phenomenology of the model. We summarise our work in section~\ref{sec:conclusions} and we detail in appendix~\ref{app:rdfit} the fit of the relic density that we have implemented. This exhibits the interplay between the variables of the model to a parametrisable curve, allowing for the prediction of relevant couplings according to the mass of the DM and mediator while accounting for co-annihilations.

\section{Heavy dark matter  with a $t$-channel mediator} \label{sec:theory}
We consider a simplified model description of the top-philic DM model \cite{Baek:2016lnv,Baek:2017ykw,Colucci:2018vxz,Colucci:2018qml} in which the SM singlet scalar DM candidate $S$ couples to the SM through a contact interaction with the SM Higgs doublet $\phi$, and a Yukawa-type interaction with the top quark and a vector-like top partner $T$. This vector-like partner has the same SM quantum numbers as the right-handed top quark, and we enforce that the masses of the new states satisfy $m_S \leq m_T$. The model has a discrete $\mathbb{Z}_2$ symmetry, under which $S$ and $T$ are odd while all SM particles are even, thus guaranteeing the stability of $S$. Following Refs.~\cite{Colucci:2018vxz,Colucci:2018qml}, we complement the SM with the Lagrangian
\begin{equation}
\mathcal{L} = i\bar{T}\slashed{D}T  -m_T\bar{T}T +  \frac{1}{2}\partial_\mu S \partial^{\mu}S - \frac{1}{2}m_S^2S^2 + \left[\tilde{y}_tS\bar{T}P_Rt + h.c.\right] + \frac{1}{2}\lambda S^2 \phi^\dagger\phi + \frac{C}{\Lambda} SS t\bar{t}.
\label{eq:CDMlag1}
\end{equation}
In our notation, $\tilde y_t$ stands for the $STt$ Yukawa coupling strength and $\lambda$ for the strength of the Higgs portal to the dark sector. The final term in this Lagrangian is a dimension-five operator linking the DM particle $S$ to the top sector via an effective contact interaction. The unknown coefficient of the contact term operator cannot be determined within the  effective theory, and we parametrise it as $C/\Lambda$ on dimensional grounds, where $C$ is a dimensionless coefficient and $\Lambda$ parametrises the scale at which the model is embedded into a more fundamental theory.\footnote{As one example, if $S$ and $T$ are bound states in an underlying composite model with a compositeness scale $\tilde{\Lambda} \sim 1$~TeV, an operator of the form $(c'/\tilde{\Lambda}^2) S \partial_S \bar{t}\gamma^\mu t$ would yield a contribution to the contact term with $C/\Lambda = c' m_t/\tilde{\Lambda}^2 $. For $c'\sim 1$, this gives $C/\Lambda \sim 0.2 \mbox{ TeV}^{-1}$.} In order to solely focus on the top-philic nature of the model, we assume that the DM coupling $\lambda$ to the Higgs field vanishes, departures from this hypothesis being discussed in  Ref.~\cite{Baek:2016lnv}. Thus, the effective Lagrangian~(\ref{eq:CDMlag1}) yields a four-dimensional parameter space  with two masses, $m_S$ and $m_T$, and two couplings, $\tilde{y}_t$ and $C/\Lambda$. In this work we will parametrise the mass-splitting between $S$ and $T$ by introducing the dimensionless quantity $r = m_T/m_S-1$.\

The DM and collider phenomenology of this top-philic DM model but without additional contact interaction has been discussed in Refs.~\cite{Baek:2016lnv,Baek:2017ykw,Colucci:2018vxz}. The study in Ref.~\cite{Colucci:2018vxz}  demonstrated the importance of QCD radiative corrections, which have a major impact on the parameter space which reproduces the observed DM relic density, as well as on direct and indirect detection prospects. This shows that the DM phenomenology of the model is potentially sensitive to {\it a priori} suppressed corrections, and motivates studying the impact of the contact interaction parametrised by  $C/\Lambda$. 

\begin{figure}
	\centering
	\includegraphics[width=0.18\textwidth]{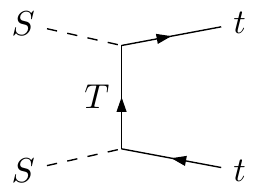}
	\hfill
	\includegraphics[width=0.18\textwidth]{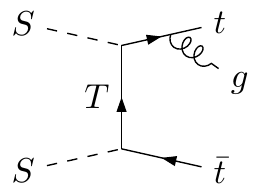}
	\hfill
	\includegraphics[width=0.18\textwidth]{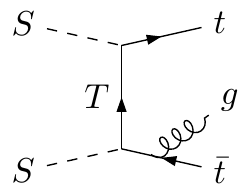}
	\hfill
	\includegraphics[width=0.18\textwidth]{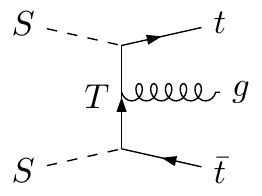}
	\hfill
	\includegraphics[width=0.18\textwidth]{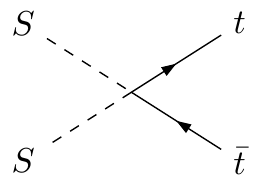}
	\caption{Representative Feynman diagrams of processes which contribute to the annihilation cross section of top-philic DM. Illustrative LO (left-most),  QCD NLO (middle) and contact term (right) contributions are displayed, where the second and third diagrams are referred to as final state radiation (FSR) and the fourth diagram is virtual internal bremsstrahlung (VIB).}
	\label{fig:feyn}
\end{figure}

Diagrams relevant for the calculation of the relic density through the $SS\rightarrow t\bar{t}$ channel are shown in figure~\ref{fig:feyn}, where the first diagram contributes to the leading-order (LO) cross section and the next three to its next-to-leading-order (NLO) corrections in the strong coupling $\alpha_s$. To these diagrams we add the process which emerges as a result of the dimension-five contact term (last diagram in the figure). Following Ref.~\cite{Colucci:2018qml}, the NLO annihilation cross section is well approximated by
\begin{equation}
\sigma v_{NLO} \approx \sigma v_{t\bar{t}} + \sigma v^{(0)}_{VIB},
\label{eq:ov}
\end{equation}
where
\begin{equation}
\sigma v_{t\bar{t}} = \frac{\tilde{y}_t^4 N_c}{4\pi m_S^3}\frac{m_t^2(m_S^2 - m_t^2)^{3/2}}{(m_S^2 + m_T^2 - m_t^2)^2}
\end{equation}
and 
\begin{equation}
\begin{split}
\sigma v^{(0)}_{VIB} = \frac{N_c \tilde{y}_t^4}{8\pi m_S^2}\frac{\alpha_S C_F}{\pi}& \biggl[\left( (r+1)^2+1\right)\left(\frac{\pi^2}{6}-\log^2\frac{1+(r+1)^2}{2(r+1)^2} - 2\text{Li}_2\left(\frac{1+(r+1)^2}{2(r+1)^2}\right)\right) \\
&+ \frac{4(r+1)^2+3}{(r+1)^2+1}+\frac{4(r+1)^{2}-3(r+1)^2-1}{2(r+1)^2}\log\frac{(r+1)^2-1}{(r+1)^2+1}\biggr],
\end{split}
\end{equation}
with $r = m_T/m_S-1$, $N_c=3$ and $C_F = 4/3$.

The $SStt$ contact interaction yields an additional contribution to the thermally averaged cross section whose leading component reads
\begin{equation}
\langle \sigma v\rangle_{SStt} =  \left(\frac{C}{\Lambda}\right)^2\frac{N_c}{4\pi}\left(1-\frac{m_t	^2}{m_S^2}\right)^{3/2}+~\mathcal{O}(v^2).
\label{eq:dim5ov}
\end{equation}
In principle, the relic density contains dimension-four contributions, dimension-five contributions, and their interference. In the relevant regions of the parameter space, the interference is however found to account for at most 1–2\%. It is therefore neglected in the following. We determine the relic density numerically (and use these results for the DM and collider phenomenology analysis in sections~\ref{sec:directdetec}, \ref{sec:inddec} and \ref{sec:collider}) and present the results first, before interpreting them through a semi-analytic description which is obtained by a fit to the numerically obtained solutions. The relic density is calculated by including both the NLO and $SStt$ contributions to the annihilation cross section. This allows for the determination of the regions of the parameter space $(m_S, m_T, \tilde{y}_t, C/\Lambda)$ in which the relic of the DM candidate matches experimental data. In order to estimate the DM relic density including not only the NLO effects but also all relevant annihilation and co-annihilation channels, we employ the {\sc MicrOMEGAs}~\cite{Belanger:2018ccd} framework, for which we generate a {\sc CalcHEP}~\cite{Belyaev:2012qa} model file through its interface to {\sc FeynRules}~\cite{Alloul:2013bka,Christensen:2009jx}. For the parameter scan we vary the mass $m_S$ ($m_T$) between $200~{\rm GeV}$ and $3000~{\rm GeV}$ ($3500~{\rm GeV}$), and allow the Yukawa coupling $\tilde{y}_t$ to lie within the range $[10^{-4},6]$. For the $C/\Lambda$ coupling we then scan over the interval  $[10^{-3}, 10^{-5}]~{\rm GeV}^{-1}$.

\begin{figure}
	\centering
	\includegraphics[width=0.7\textwidth]{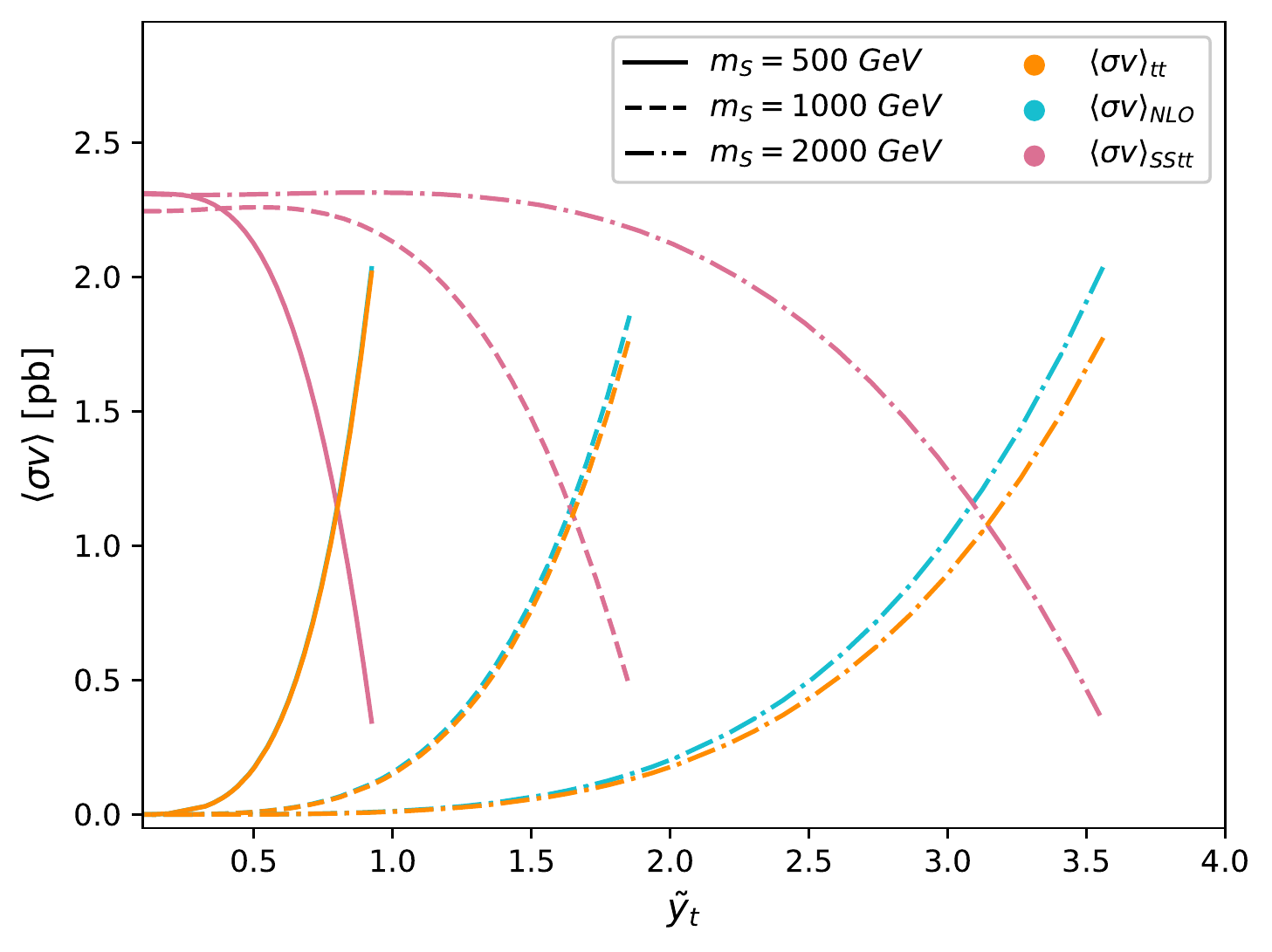}
	\caption{A comparison of the impact of the different contributions to the thermally-averaged DM annihilation cross section for setups that yield the observed DM relic density. We vary the Yukawa coupling $\tilde{y}_t$, have chosen three DM benchmark masses $m_S$ and fixed $m_T$ so that $r=0.6$. The last parameter ($C/\Lambda$) has been fixed to match Planck's results. This figure shows the interplay of the contributions from LO ($\langle\sigma v \rangle _{tt}$, orange), NLO ($\langle\sigma v \rangle _{NLO}$, blue), and contact interaction ($\langle\sigma v \rangle _{SStt}$, magenta) diagrams.}
	\label{fig:comparexsecs}
\end{figure}

The relative impact of the LO, NLO and contact operators to the cross sections is illustrated in figure~\ref{fig:comparexsecs}, for a small sample of benchmark scenarios. We have considered three DM masses of $m_S=500$, 1000 and 2000 GeV and fixed the heavy-top mass $m_T$ through $r=0.6$. We have then varied the Yukawa coupling $\tilde y_t$ from 0 to 4, and updated the $C/\Lambda$ value to match the observed relic density. The figure exhibits the interplay between the different terms that contribute to the total annihilation cross section $\langle \sigma v \rangle = \langle \sigma v\rangle_{NLO} +\langle  \sigma v\rangle_{SStt}$, where the separate contributions sum to the thermally averaged cross section required to yield the observed DM relic density. As can be seen, small $\tilde{y}_t$ yields a small NLO cross section which is compensated for by the  contact operator contribution. Also shown in the figure is the LO contribution, which is accounted for in the NLO result~(\ref{eq:ov}), where the VIB processes become increasingly relevant for higher masses $m_S$.

This figure shows the importance of the dimension-five contribution from the contact operator to the total annihilation cross section in the regime of small Yukawa coupling, and motivates further investigation into potential modifications to the phenomenology and direct and indirect detection constraints. The strength of the contact interaction is of order $C/\Lambda\simeq 0.2 \mbox{ TeV}^{-1}$ (or less, if the cross section is dominated by the NLO contribution) and in the following we determine a semi-analytic approximation for the value of $C/\Lambda$ required to reproduce the observed DM relic density.

The DM relic density obeys the Boltzmann equation\footnote{For the moment, we neglect co-annihilation effects.}
\begin{equation}
\frac{{\rm d}n}{{\rm d}t} = -3 H n - \langle \sigma v\rangle \left(n^2-n_{eq}^2\right),
\label{eq:boltzmann}
\end{equation}
where $H$ is the Hubble constant, $n$ denotes the  number-density of $S$, $n_{eq}$ is the equilibrium number density, and  $ \langle \sigma v\rangle $ is the thermally averaged annihilation cross section. An approximate solution to the Boltzmann equation is given by~\cite{Kong:2005hn} 
\begin{equation}
\Omega_{DM}h^2 \approx \frac{1.04 \times 10^9}{M_{Pl}}\frac{x_F}{\sqrt{g_*(x_F)}}\frac{1}{a + 3b/x_F}, 
\label{eq:fullom}
\end{equation}
where $M_{Pl}$ is the Planck mass, $g_*$ is the total number of effectively massless degrees of freedom, and $x_F$ is the ratio of the dark matter mass to the freeze-out temperature. The $a$ and $b$ coefficients are functions of the masses and are defined from the non-relativistic annihilation cross section, which we can expand as
\begin{equation}
\langle \sigma v \rangle = a + b \langle v^2 \rangle.
\end{equation} 
For the currently considered model, the thermally averaged cross section is a sum of equations~(\ref{eq:ov}) and~(\ref{eq:dim5ov}),
\begin{equation}
\langle\sigma v\rangle = \langle \sigma v\rangle_{NLO} +\langle {\sigma} v\rangle_{SStt}.
\label{eq:annihlxs}
\end{equation}
The dependence on the couplings $\tilde{y}_t$ and $C/\Lambda$ can be factored out, so that
\begin{equation}
\langle \sigma v\rangle_{NLO} = \tilde{y}^4_t B(m_S,m_T)
\qquad\text{and}\qquad
\langle \sigma v\rangle_{SStt} =\left(C/\Lambda\right)^2A(m_S),
\end{equation}
with $A(m_S)$ being a function of the DM mass and $B(m_S,m_T)$ being a function of both the DM and the mediator masses. Therefore, using that $\Omega_{DM}h^2 \sim 1/ \langle \sigma v\rangle$~\cite{Lisanti:2016jxe}, solving for $C/\Lambda$ yields
\begin{equation}
\frac{C}{\Lambda} \approx \frac{1}{ \sqrt{A(m_S)}}\sqrt{b^\prime(x_F, g_*(x_F)) - B(m_S,m_T)~ \tilde{y}_t^4},
\label{eq:funcfit}
\end{equation}
with 
\begin{equation}
\begin{split}
A(m_S)  &=\frac{\Lambda^2\langle \sigma v\rangle_{SStt}}{C^2} =  \frac{N_c}{4\pi}\left(1-\frac{m_t	^2}{m_S^2}\right)^{3/2},\\
B(m_S,m_T) &=  \frac{\sigma v_{q\bar{q}} + \sigma v^{(0)}_{VIB}}{\tilde{y}_t^4}\\
&  = \frac{N_c}{4\pi m_S^2}\bigg(\frac{m_t^2(m_S^2 - m_t^2)^{3/2}}{m_S(m_S^2 + m_T^2 - m_t^2)^2} \\
&+ \frac{\alpha_S C_F}{2\pi}\bigg[((r+1)^2+1)\left(\frac{\pi^2}{6}-\log^2\frac{1+(r+1)^2}{2(r+1)^2} - 2\text{Li}_2\left(\frac{1+(r+1)^2}{2(r+1)^2}\right)\right)  \\
& + \frac{4(r+1)^2+3}{(r+1)^2+1} +\frac{4(r+1)^{2}-3(r+1)^2-1}{2(r+1)^2}\log\frac{(r+1)^2-1}{(r+1)^2+1}\bigg]\bigg),\\
b^\prime(x_F, g_*(x_F)) & =  \left(7.2 \times 10^{-10}~ \mbox{GeV}^{-2} \right) \frac{x_F}{\sqrt{g_*(x_F)}}.
\end{split}
\label{eq:AB}
\end{equation}
The $b^\prime(x_F, g_*(x_F))$ function is then determined from a fit to the numerical result.

For the result in eq.~(\ref{eq:funcfit}) we neglected co-annihilation effects. They play a role when the DM candidate $S$ and the mediator $T$ are nearly mass degenerate, which is the case in parts of the parameter space considered. The co-annihilation effects, too, can be treated semi-analytically, and we refer the interested reader to Appendix \ref{app:rdfit}. Here, we just quote the final result that generalises eq.~\eqref{eq:funcfit},
\begin{equation}
\frac{C}{\Lambda} \approx f(m_S,m_T,\tilde{y}_t) = \frac{1}{\sqrt{A(m_S)}} 
\sqrt{b^\prime - B(m_S,m_T) \left(\tilde{y}_t- \alpha \left[ \beta\gamma^{\frac{m_S}{\Lambda}}\right]^r\right)^4},
\label{eq:full}
\end{equation}
where $A(M_s)$ and $B(m_S, m_T)$ are given in eq.~(\ref{eq:AB}), $r = m_S/m_T-1$, and the coefficient $b^\prime(x_F, g_*(x_F))$ is determined by a fit to the numerical results to be $b^\prime=6.0 \times 10^{-9} \pm 0.2 \times 10^{-9}~ {\rm GeV}^{-2}$. The other coefficients which parametrise the co-annihilation effects are fitted to $(\alpha,\beta,\gamma)=(0.4, 1.9\times 10^{-4},6.2\times 10^8)$ for $m_S\leq 1.2~{\rm TeV}$, and $(\alpha,\beta,\gamma)=(0.7, 3.0\times 10^{-3},1.8\times 10^4)$ for $m_S > 1.2~{\rm TeV}$. The parameter $\gamma$ has been raised to a dimensionless ratio featuring $\Lambda = 3.5~{\rm TeV}$, which is the maximum value for $m_T$ used in the scan. Such a value hence provides an indicative scale of the effective model (such as the limit of validity of the theory or the scale of compositeness).

\section{Direct detection prospects and bounds} \label{sec:directdetec}
As the DM candidate in our model interacts with ordinary matter, its properties can be probed at direct DM detection experiments. In such experiments, the collision of $S$ with a nucleus of the detector material can leave hints to be observed through the recoil energy of the nucleus. The rate at which this occurs is related to the nucleon-DM cross section as predicted in our model. As the DM scalar $S$ only interacts with the Higgs boson (although we neglect such interactions) and the top quark, it does not have any tree-level interactions with valence quarks of the nuclei, and DM-gluon scattering at the loop level is the dominant contribution to DM-nucleus scattering.

\begin{figure}
	\centering
	\includegraphics[width=0.2\textwidth]{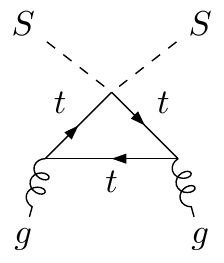}
	\caption{Additional diagram to be considered in the evaluation of the $S$ coupling to a nucleus via the constituting gluons of the latter, once the model Lagrangian includes an $SStt$ dimension-five operator.}
	\label{fig:addamps}
\end{figure}

In evaluating the scattering cross section with a nucleon we follow Ref.~\cite{Hisano:2015bma}, matching the effective theory describing the DM-nucleon interactions to the full theory through higher-dimensional operators. In the case of scalar DM, only the spin-independent cross section is relevant. As the low velocity of the DM leads to a small momentum transfer~\cite{Belanger:2020gnr}, the interaction between the DM and nucleons can be described by the following effective Lagrangian
\begin{equation}
\mathcal{L} =  C_S^g(m_T)\ \mathcal{O}^g_S =  C_S^g(m_T)\ \frac{\alpha_s}{\pi}S^2 G^{\mu\nu}G_{\mu\nu} ,
\end{equation}
where the effective operator is defined at the mass scale of the mediator. The Wilson-coefficient $C_S^g$ receives contributions $f^{d4}$ resulting from the renormalisable part of the Lagrangian~(\ref{eq:CDMlag1}), as well as contributions $f^{d5}$ originating from the dimension-five contact interaction examined in this work,
\begin{equation}
C_S^g = \frac{\tilde{y}^{2}_t}{8} f^{d4}(m_S, m_t, m_T) + f^{d5}(m_t).
\label{eq:cgorig}
\end{equation}
The full expression for $f^{d4}(m_S, m_t, m_T)$ has been determined in Refs.~\cite{Hisano:2015bma,Colucci:2018vxz} (and are in particular collected in the appendix of Ref.~\cite{Hisano:2015bma}). The new contribution $f^{d5}$  arises from the addition of the $SStt$ coupling to the model Lagrangian. It leads to an additional diagram  to the total amplitude for the DM interaction with the nucleus, displayed in figure~\ref{fig:addamps}. The effective coupling is written as
\begin{equation}
f^{d5}(m_t) = - \frac{C}{\Lambda}\int \frac{{\rm d}^4p}{(2\pi)^4}\text{Tr}[i S(p)]\vert_{GG} = -\frac{C}{\Lambda} \frac{1}{m_t},
\end{equation}
where ``$\vert_{GG}$'' indicates that the coefficient for the terms proportional to $G_{\mu\nu}^a G^{a\mu\nu}$ have been extracted from the quark propagator in the gluon background in the limit of zero gluon momentum~\cite{Hisano:2010ct}
\begin{equation}
i S(p) = \int {\rm d}^4x e^{ipx} \langle T \{ \psi(x) \overline{\psi}(0) \} \rangle = \frac{1}{m_t}.
\end{equation}

With the Wilson coefficient determined, the spin-independent DM-nuclear cross section $\sigma_A$ (for $n_p$ protons and $n_n$ neutrons in a nucleus of mass $m_A$) and DM-proton cross section $\sigma_p$ read
\begin{equation}
\begin{split}
&\sigma_A = \frac{1}{\pi}\left(\frac{m_A}{m_S + m_A}\right)^2|n_p f_p + n_n f_n|^2\\
&\sigma_p = \frac{1}{\pi}\left(\frac{m_p}{m_S + m_p}\right)^2f_p^2.
\end{split}
\label{eq:ddxsec}
\end{equation}
For a nucleon $N$ we moreover have
\begin{equation}
f_N/m_N = -\frac{8}{9}C^g_S f^{(N)}_{T_G}, \quad f_{T_G}^{(N)} = 1 - \sum_{q = u, d, s} f_{T_q}^{(N)}
\end{equation}
with the quark mass fractions $f_{T_q}^{(N)}$ being given in table~1 of Ref.~\cite{Hisano:2015bma}. In this notation, the dependence on the strong coupling constant is implicitly absorbed in the definition of $f_N$.\

The scattering cross sections in eq.~(\ref{eq:ddxsec}) are proportional to $\left(C^g_S\right)^2$, and $C^g_S$ receives a positive contribution proportional to $\tilde{y}^2_t$ from the renormalisable interactions in the TeV-scale theory. In contrast, the contact interaction contribution is proportional to $-C/\Lambda$. The sign of $C$ is not fixed at the EFT level, such that the contact interaction can either increase or decrease the direct detection cross section, depending on the sign of $C/\Lambda$. While the sign of $C/\Lambda$ plays an important role here, we recall that it was not relevant in the determination of the DM relic density.

\begin{figure}
	\centering
	\includegraphics[width=0.49\textwidth]{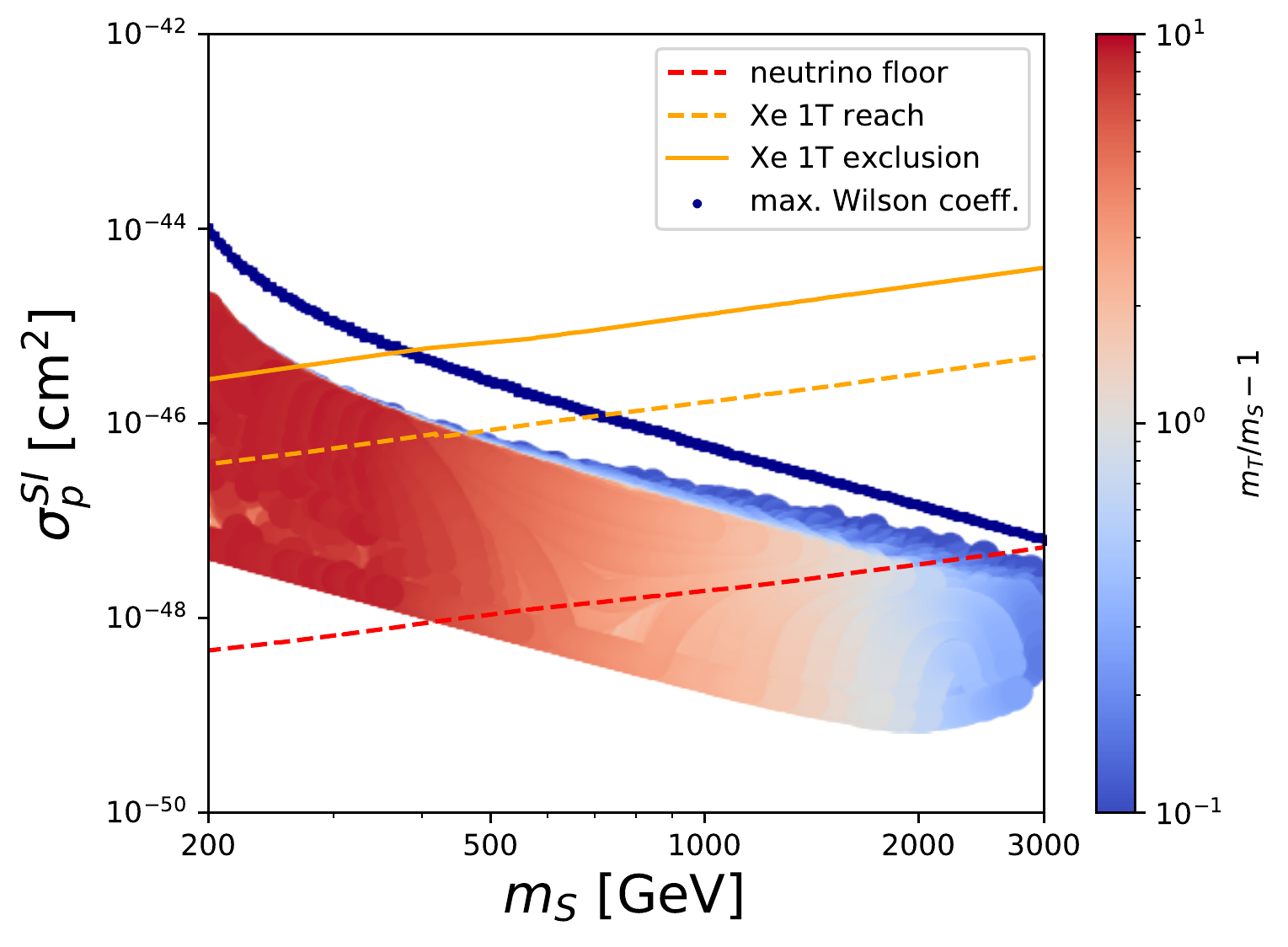}
	\includegraphics[width=0.49\textwidth]{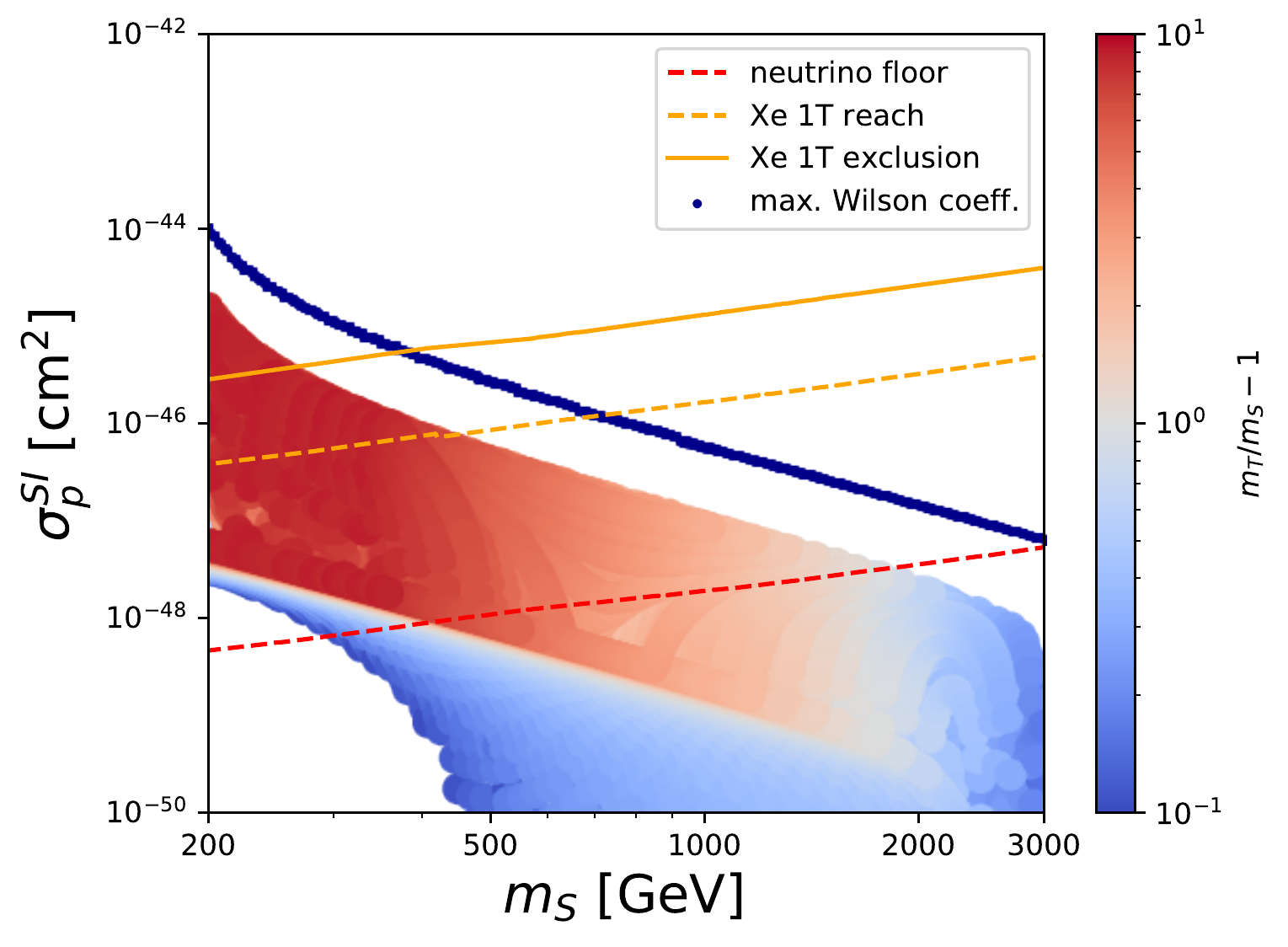}
	\caption{The DM-proton cross section $\sigma_p$ for positive (left) and negative (right) $C/\Lambda$ values, and for the maximum value of the Yukawa coupling $\tilde y_t$ so that there exists a combination of the other model parameters that gives the correct relic density. The $r=m_T/m_S-1$ value (shown through the red-and-blue colour scheme) is enforced to lie in a [0.1,9] interval and is derived from the value of the other model parameters and the relic density. The bottom limit of 0.1 ensures that we are in a well-understood regime of spectrum compression, and the upper limit is chosen such that the vector-like resonance and the scalar DM are not too different in mass. Additionally, we display the cross section obtained through use of the maximum possible Wilson coefficient $|C|$, which is displayed in dark blue. In those scenarios the dependence of the relic density on $\tilde y_t$ is negligible with respect to the contribution from the $SStt$ interactions. Finally, the red dashed line represents the neutrino floor~\cite{Billard:2013qya}, the orange dashed line indicates the XENON 1T reach~\cite{Aprile:2015uzo}, and the red solid line shows the 90\% confidence exclusion of the XENON 1T experiment~\cite{Aprile:2017iyp}.}
	\label{fig:DMprotonfull}
\end{figure}

In figure~\ref{fig:DMprotonfull} we present the DM-proton cross section $\sigma_p$ across the full considered $m_S$ mass range. Recalling that we consider the parameter ranges $C/\Lambda~\in~[10^{-5},~10^{-3}]~{\rm GeV}^{-1}$ and $\tilde{y}_t~\in~[10^{-4},~6]$, we expect non-zero contributions from both the NLO and dimension-five processes in determining the full cross section displayed in the figure. Here and for the remainder of the paper, we consider $r=m_T/m_S-1~\in~[0.1,~9]$, where the upper limit ensures that $S$ and $T$ are separated in mass by at most an order of magnitude. This is motivated by the idea that both emerge as resonances in a composite Higgs model. The lower limit is chosen as the compression scale at which co-annihilation effects start to become apparent (see appendix~\ref{app:rdfit} for further details). While greater compression can certainly be modelled by the tools at hand, we choose to stay far from co-annihilation effects as we do not fully account for them. \

At each DM mass, we first single out the ensemble of scanned scenarios featuring the right relic density, and then plot the $\sigma_p$ values associated with each scenario from a selected subset. In this subset of models, the absolute value of the Wilson coefficient $|C/\Lambda|$ is minimum, and the value of the Yukawa coupling $\tilde y_t$ is thus maximum. The corresponding value for the compression factor $r$ is also depicted, through the red-and-blue colour scheme. Moreover, we superimpose to our predictions constraints and the projected reach from the XENON experiment, as well as the neutrino floor.  Conversely, the dark blue line includes the results obtained when selecting a subset of scenarios for which the Wilson coefficient is maximum, and the Yukawa coupling $\tilde y_t$ is thus at a minimum. This figure demonstrates that many viable models from the relic density standpoint lie above the neutrino floor, and that some models, particularly in the low DM mass region, are within the reach of future upgrades of the XENON experiment. Those are thus in principle testable in the future. In the figure, the difference in behaviour for positive and negative $C/\Lambda$ is also highlighted.

For a positive dimension-five coupling, the viable regions of parameter space display significant overlap across $r = m_T/m_S -1$ values. This is in contrast to the case of a negative dimension-five coupling, where the resulting parameter space is more spread out, and many scenarios with lower $r$ values result in very small $\sigma^{\rm SI}_p$, which are not viable for detection. The benchmark points with $r\gtrsim 1$ values are common to both setups, {\it i.e.}\ the reddish band across the centre of the figure is unchanged for positive and negative $C$. 
We recall that such a band corresponds to low $|C/\Lambda|$ and large $\tilde{y}_t$, where $C$ is non-zero and the Yukawa coupling has taken over entirely in the calculation of the relic density. The implications for direct detection are markedly different from the relic density calculations for low $|C/\Lambda|$; in the case of the relic density, the minimum value for $|C/\Lambda|$ was chosen to reflect the case where the dimension-five term no longer modifies the relic density produced by the Yukawa term, producing a relic density comparable to the case $C=0$. However, the same minimum value for $|C/\Lambda|$ clearly modifies the direct detection prospects in a non-negligible way.\
As soon as $|C/\Lambda|$ increases from its minimum value, so does the DM-proton scattering cross section to finally reach the bright blue line on the figure, that corresponds to a maximum $|C/\Lambda|$ value. Here, the dimension-five addition completely dominates the DM-proton cross section and takes over from the Yukawa term, leading to an unchanged cross section under the $C$ sign flip. Even for `small' $C/\Lambda$, as is considered here, the contribution from the dimension-five operator is still larger than the one proportional to the $\tilde y_t$ Yukawa coupling.

It is clear from the figure that the difference between positive and negative $C$ emerges for smaller values of $r$. For $r$ values smaller than about 1 (in particular for $m_S > 500$~GeV), the $C/\Lambda$ contribution is about the same order of magnitude as the $\tilde y_t$ contribution, such that when $C/\Lambda$ is positive and the contributions have the same sign, they add constructively and make a larger cross section. When $C/\Lambda$ is negative, they add destructively and $\sigma_p$ may be pushed down. We find that, for a representative point $r~=~0.1$, heavy and light DM behave very differently; for light scalars the addition of the dimension-five greatly increases the cross section, but for positive and negative $C$ there is no change. In the case of heavy scalars and $r~=~0.1$, a sign flip in $C$ leads to destructive interference. This contrasts with $r \geq 1$ configurations in which the cross section is unchanged under the flip of the sign, no matter $m_S$.

The limiting values for the intersection with the neutrino floor, the XENON 1T reach, and the 90\% confidence exclusion of the XENON 1T, are given in table~\ref{tab:intersectdd}. In this table we give the mass $m_S$ at which the relevant exclusion intersects the distribution of phenomenologically viable scenarios, for both positive and negative Wilson coefficients. It is notable that the most significant difference arises in the case of the neutrino floor. There is a much larger number of scenarios below it, and thus potentially hard to probe, in the case of a negative Wilson coefficient. We now investigate this last property deeper.

\begin{table}
	\centering
	\renewcommand\arraystretch{1.55}
	\setlength\tabcolsep{8pt}
	\begin{tabular}{c||c|c|c|c}
		\multirow{2}{*}{Exclusion} & \multicolumn{2}{c|}{Max. $\tilde{y}_t$} & \multicolumn{2}{c}{Max. $C/\Lambda$} \\
		
		& $C > 0$ & $C<0$ & $C>0$ &  $C<0$\\
		\hline
		\hline
		Xe-1T exclusion & 248 GeV & 246 GeV &  \multicolumn{2}{c}{376 GeV} \\
		\hline
		Xe-1T reach & 405 GeV & 393 GeV & \multicolumn{2}{c}{706 GeV}  \\
		
		\hline
		$\nu$-floor & 2476 GeV & 1631 GeV& \multicolumn{2}{c}{$>$ 3000 GeV} \\
		
	\end{tabular}
	\caption{The masses $m_S$ at which the bounds in figure~\ref{fig:DMprotonfull} intersect the regions of interest.}
	\label{tab:intersectdd}
\end{table}

\begin{figure}
	\includegraphics[width=0.48\textwidth]{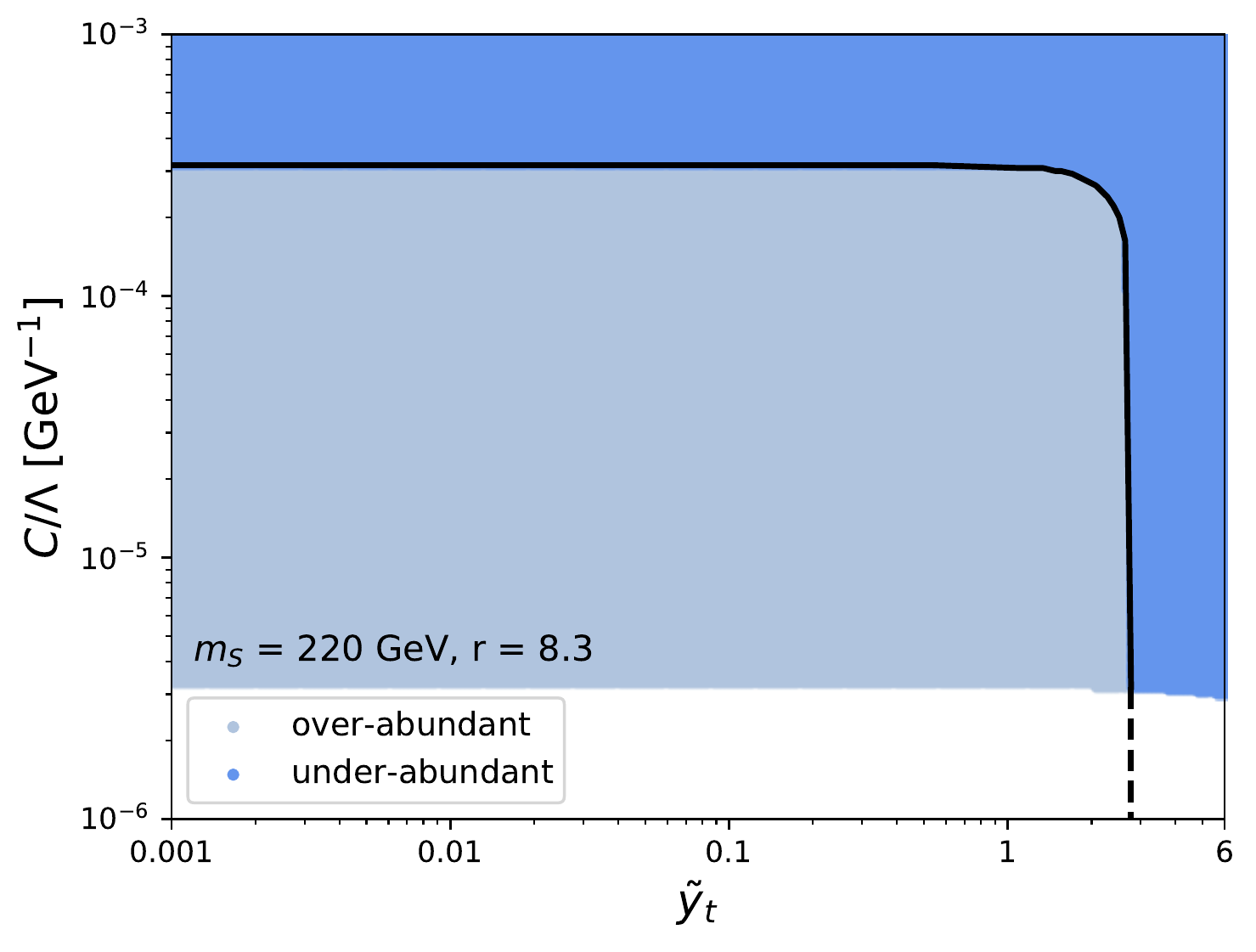}
	\hfill
	\includegraphics[width=0.48\textwidth]{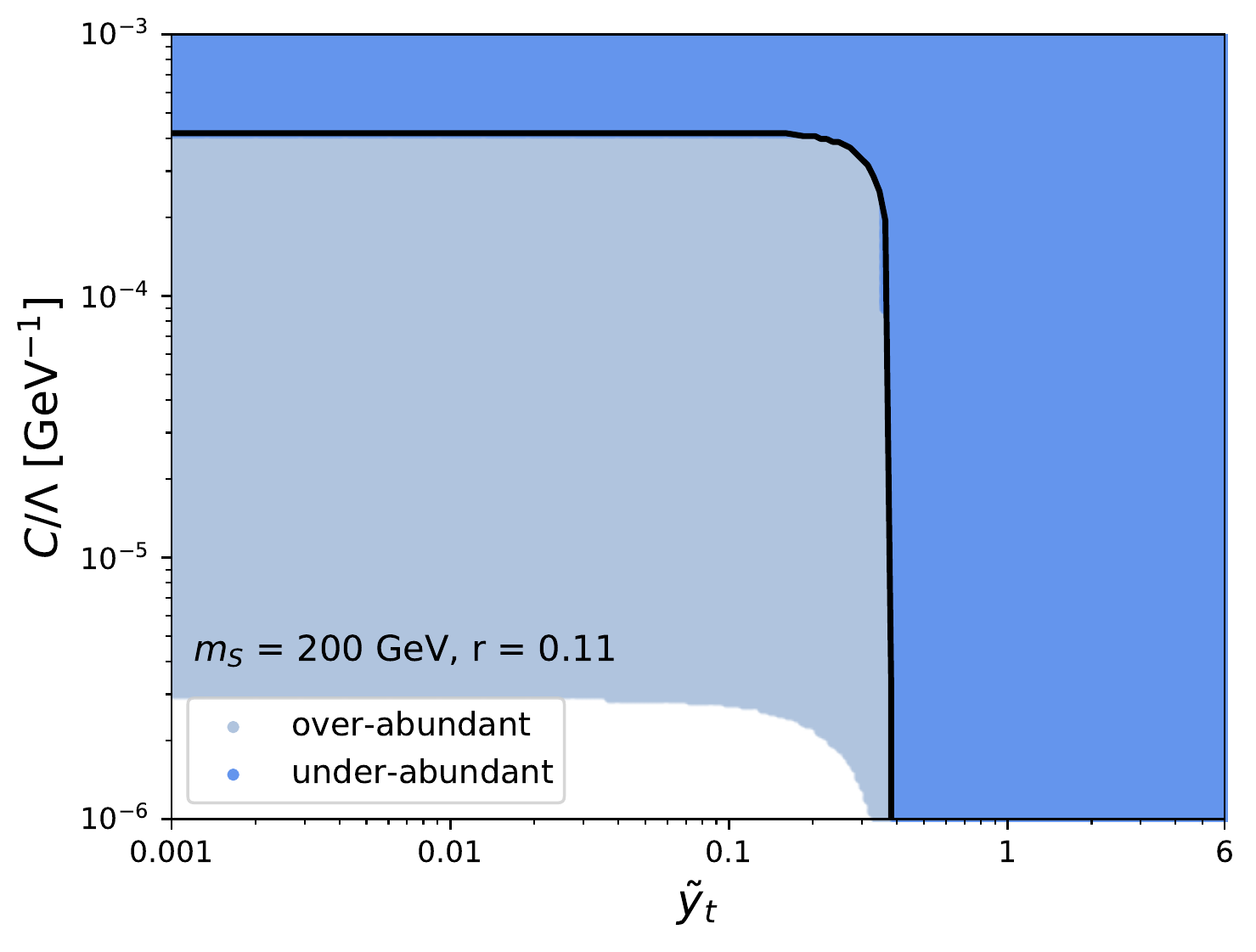}
	\hfill
	\includegraphics[width=0.48\textwidth]{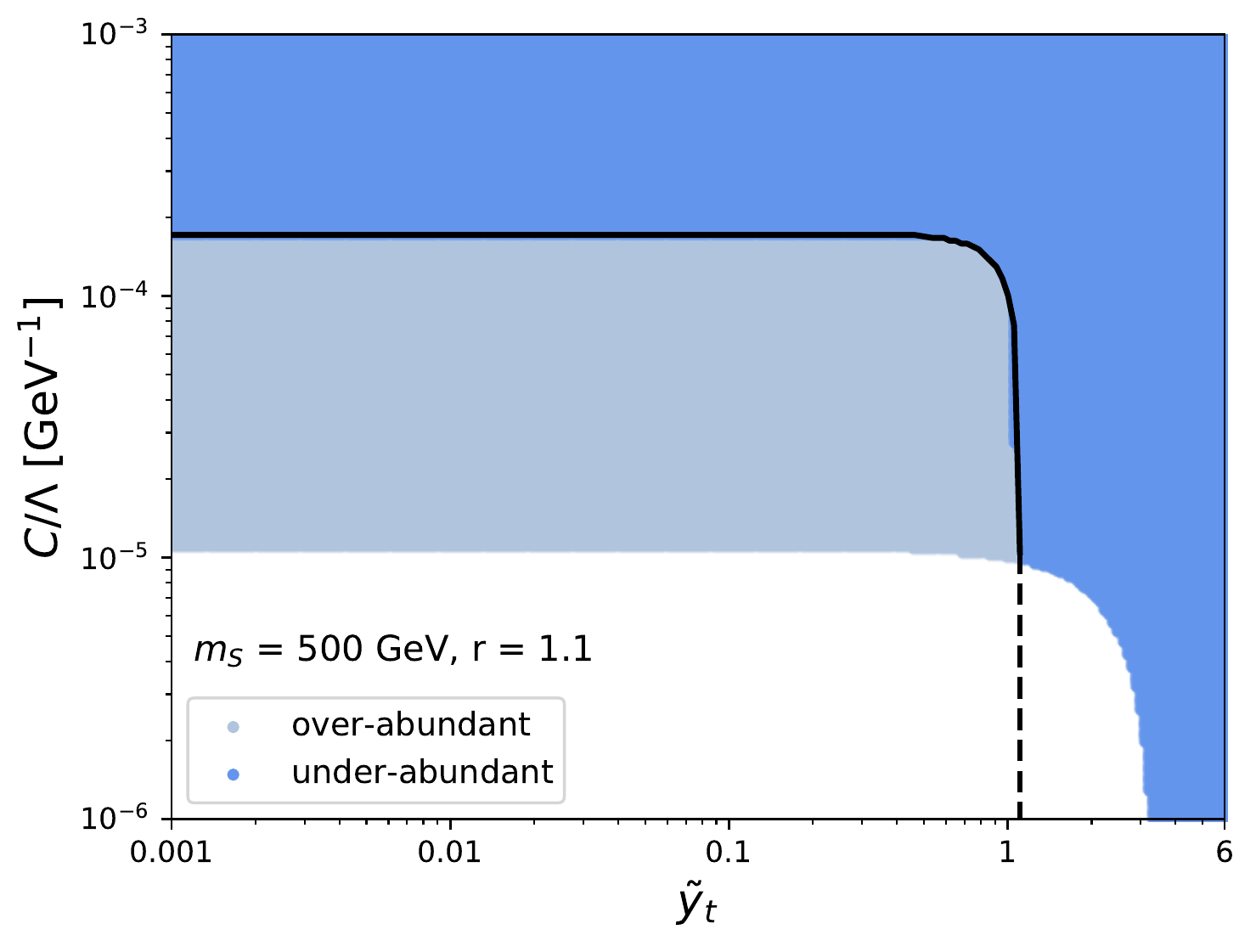}
	\hfill
	\includegraphics[width=0.48\textwidth]{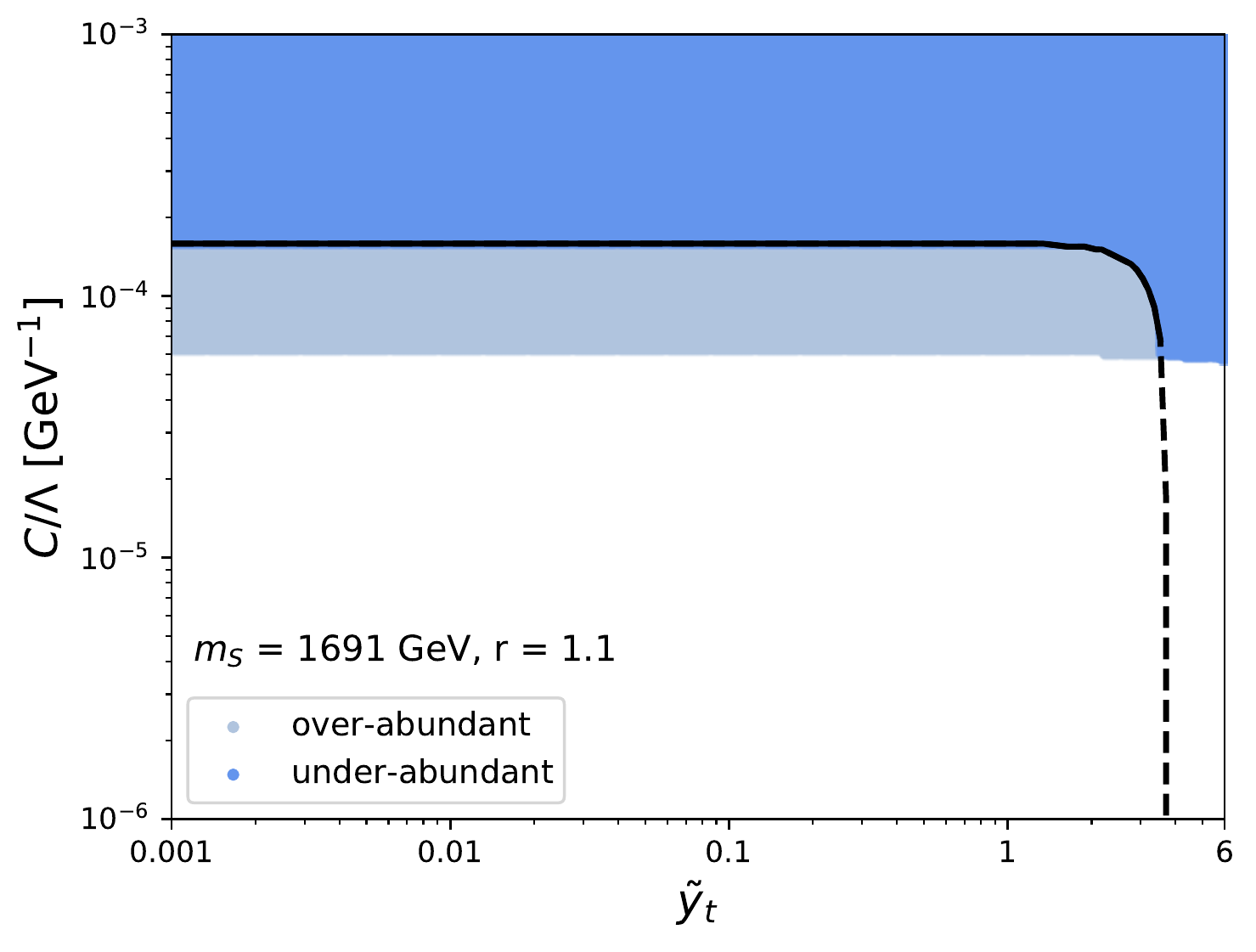}
	\hfill
	\includegraphics[width=0.48\textwidth]{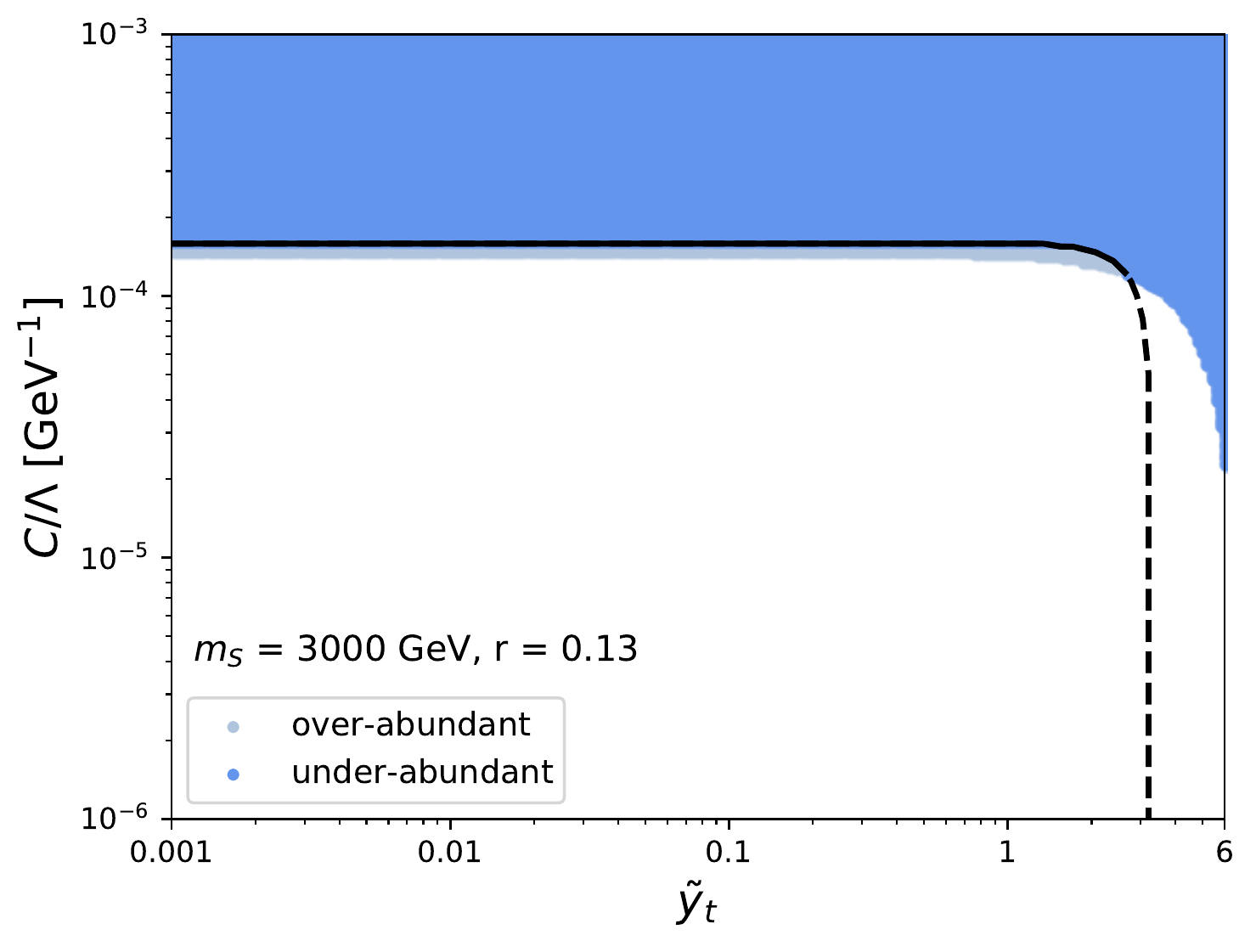}
	\hfill
	\includegraphics[width=0.48\textwidth]{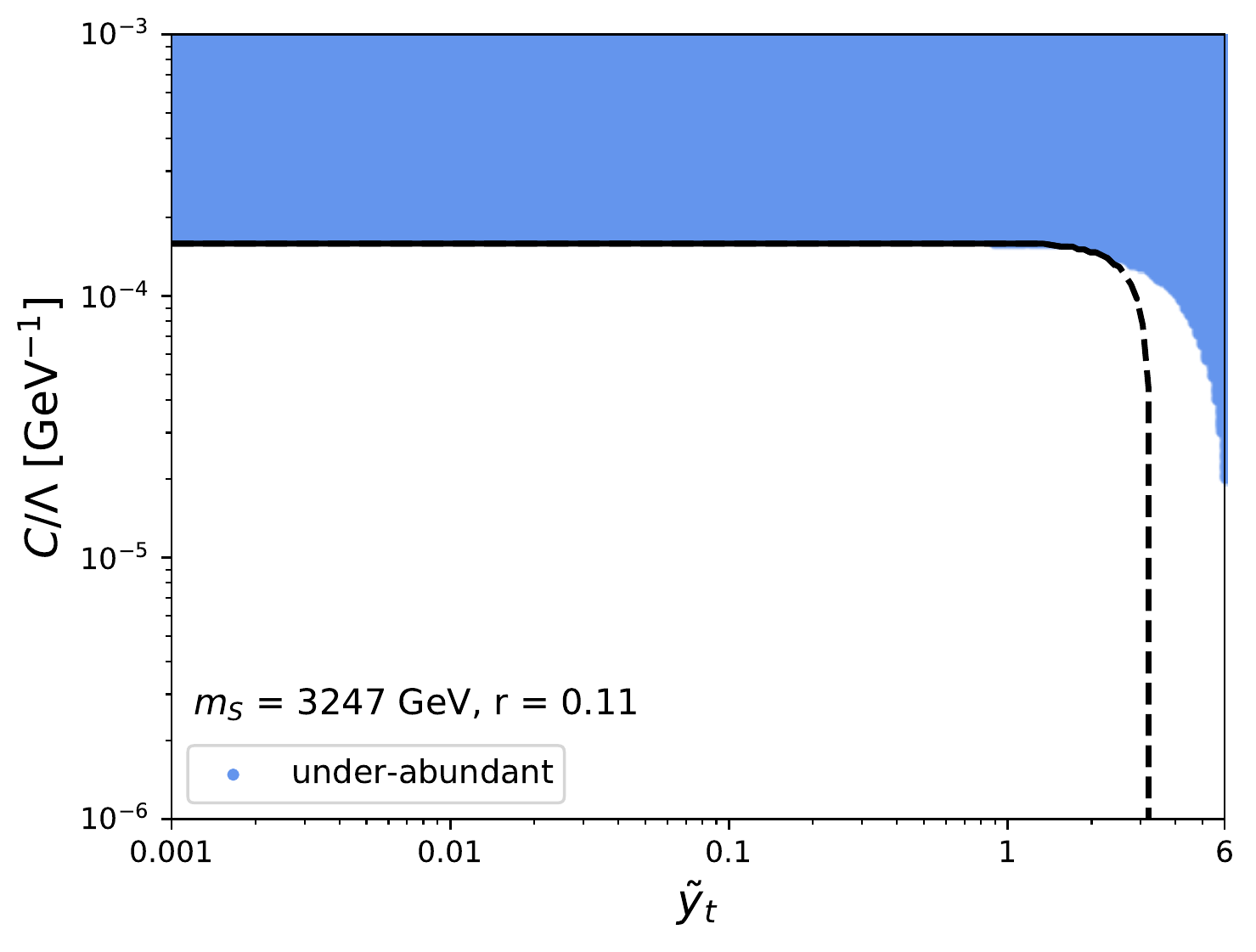}
	\caption{Six classes of benchmarks illustrating a variety of regimes. These include compressed, split, light, and heavy mass points. The blue area denotes the area in the $(C/\Lambda, \tilde{y}_t)$ plane which corresponds to a DM-proton cross section which is larger than the neutrino floor (these points do not necessarily correspond to the correct relic density). The relationship between the two parameters which leads to the correct relic density $\Omega_{DM}h^2$ = 0.1186 $\pm$ 0.0020 is shown as a black line, where the solid line corresponds to the parameter space lying above the neutrino floor, and the dashed line represents points hidden below the neutrino floor.}
	\label{fig:cyexclusion}
\end{figure}

In figure~\ref{fig:cyexclusion}, we determine the region of interest in the ($C/\Lambda, \tilde{y}_t$) parameter space for a number of illustrative benchmarks in increasing $m_S$. The shaded regions correspond to parameters where the DM-proton cross section crosses the neutrino floor, and is therefore potentially reachable by experiments. The shaded region has been coloured light blue if the region leads to a relic density where the DM is over-abundant, and darker blue if the DM is under-abundant. The black line indicates the parameters yielding the correct relic density. Given that $\Omega_{DM} h^2 \propto \langle \sigma v \rangle ^{-1}$, the area outside/above the relic function line is under-abundant and not excluded, and the area below the relic line is over-abundant and therefore excluded. The dotted black line indicates scenarios with a correct relic density, but in a region where the DM-proton cross section lies below the neutrino floor.

The $C/\Lambda$ parameter clearly plays a dominant role in pushing the DM-proton cross section across the neutrino floor in the region corresponding to the correct relic density when only relying on these two terms. We can relate the bright blue line visible in figure~\ref{fig:DMprotonfull} that corresponds the maximum $C/\Lambda$ coupling to the horizontal portion of the black lines in figure~\ref{fig:cyexclusion}. Moreover, the top plots in figure~\ref{fig:cyexclusion}, both at $m_S$ around 200~GeV, illustrate how a change in mass splitting modifies the Yukawa coupling impacting the features of the models in direct detection experiments. A stronger compression indeed reduces the Yukawa coupling value at which the models will become visible relative to the neutrino floor. \

\section{Indirect detection prospects and bounds} \label{sec:inddec}
Experiments which seek to detect DM through indirect methods aim to observe hints from DM annihilations or decays into SM particles that then reach us in the form of gamma or cosmic rays which may travel through the universe with little other interaction~\cite{Conrad:2014tla}. The flux of these SM particles depends on the annihilation cross section of the DM, the relative branching ratios of the different final-state particles produced in those annihilations, the mass $m_S$, as well as astrophysical constraints. The indirect detection of DM through cosmic rays holds the advantage that we can potentially detect DM on galactic or cosmological scales. In order to assess whether the indirect detection bounds may differ from those in previous works~\cite{Colucci:2018vxz}, we again utilise {\sc FeynRules} for the generation of UFO model files~\cite{Degrande:2011ua}, this time as inputs into {\sc MadGraph5\_aMC@NLO (MG5\_aMC)}~\cite{Alwall:2014hca}. Next, we simulate DM annihilation at close to zero velocity, using {\sc Pythia}~8~\cite{Sjostrand:2014zea} to describe parton showering and hadronisation. \

\begin{figure}
	\centering
	\includegraphics[width=0.49\textwidth]{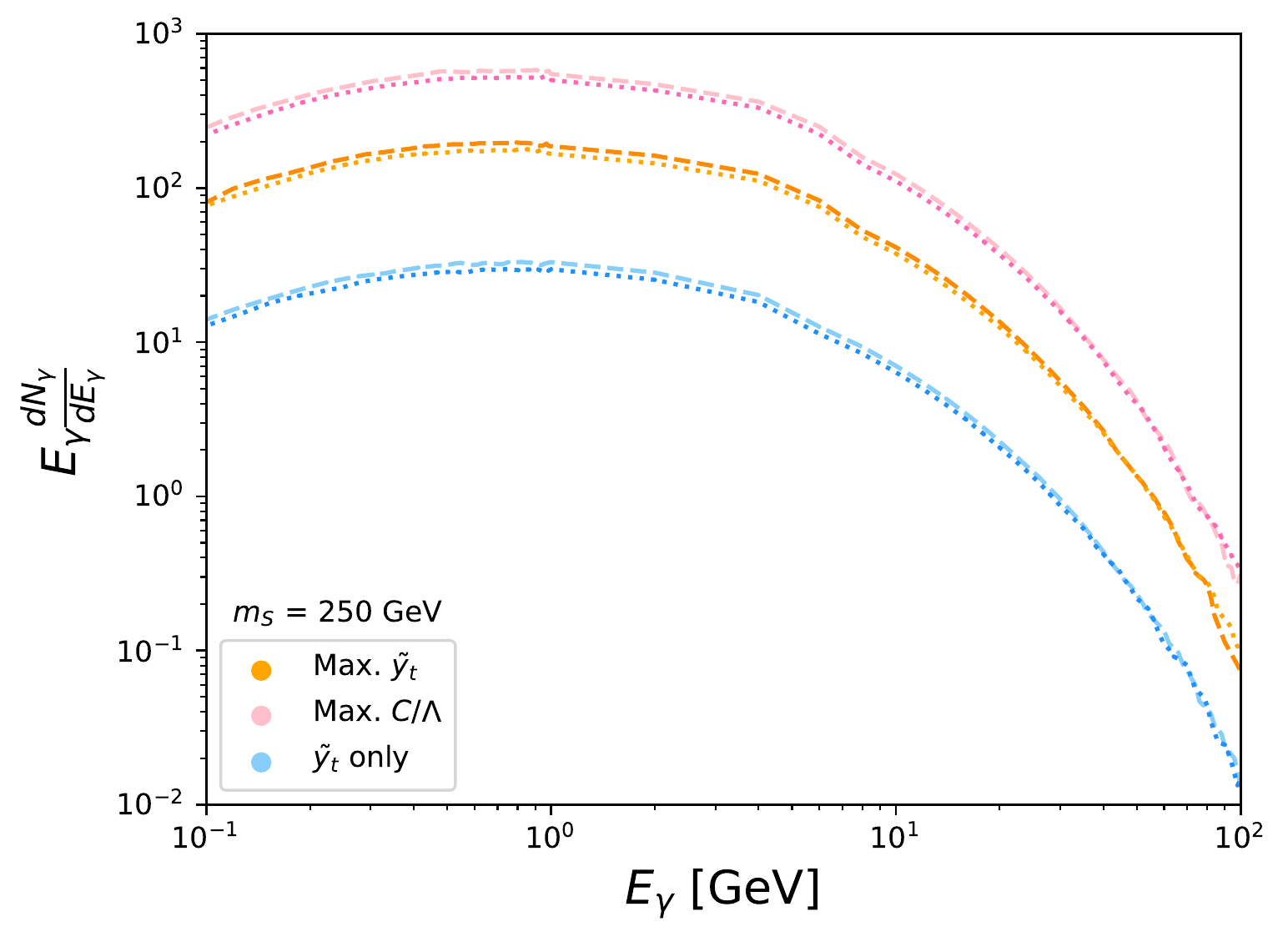}
	\includegraphics[width=0.49\textwidth]{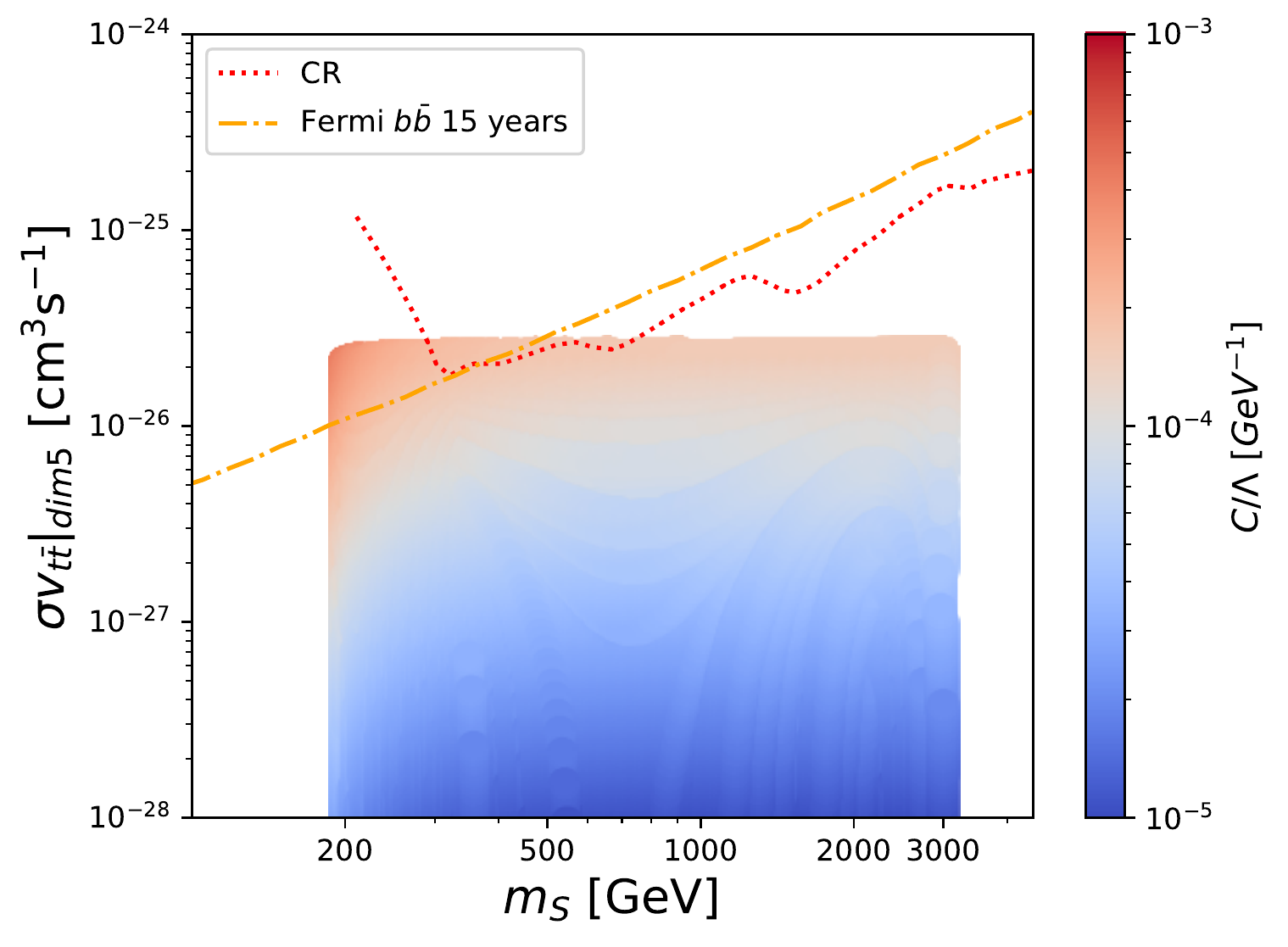}
	\caption{{\it Left} -- Photon spectra resulting from $SS\rightarrow t\bar{t}$ (dashed line) and $SS\rightarrow b\bar{b}$ (dotted line) annihilations, for a dark matter benchmark mass $m_S = 250~{\rm GeV}$. The lines correspond to parameter choices for which the DM relic density is correctly reproduced  while $C/\Lambda$ is maximal (pink), $\tilde{y}_t$ is maximal (orange), or $C/\Lambda = 0$ (blue).
		{\it Right} --  Contribution to the $SS\rightarrow tt$ annihilation cross section at zero velocity (as a function of $m_S$) due to the inclusion of the dimension-five contact interaction. The colour scale indicates the value of $C/\Lambda$ required to obtain the observed relic density.  Also shown are bounds due to cosmic rays~\cite{Cuoco:2017iax} (dotted red), and projected bounds from the Fermi-LAT dwarf spheroidal galaxy data in the $b\bar{b}$ channel~\cite{Charles:2016pgz}  (dashed orange).}
	\label{fig:egamma}
\end{figure}

In order to compare our predictions to the exclusions from experiments, we examine the gamma-ray spectrum from $b\bar{b}$ and $t\bar{t}$ DM annihilation in three cases of interest: with no dimension-five operator present, with the lowest allowed value for the dimension-five coupling in the range deduced from the relic density scan, $C/\Lambda \in [10^{-5}, 10^{-3}]~{\rm GeV}^{-1}$, and the corresponding highest Yukawa coupling for given mass setups, and finally with the highest dimension-five coupling and lowest Yukawa for the given mass setups. The results are displayed in figure~\ref{fig:egamma} (left), where initial and final state radiation are both included in the {\sc Pythia}~8 simulations. From this figure, two conclusions are evident. First, that the $t\bar{t}$ and $b\bar{b}$ final states display very similar behaviour, and so exclusion limits from DM annihilations may be rescaled to constrain $t\bar t$ final states. An estimate of the limits for the $t\bar{t}$ final state can be obtained through the rescaling of existing constraints on the $b \bar b$ final state,
\begin{equation}
\sigma v_{t\bar{t}} = \sigma v_{b\bar{b}} \frac{N_\gamma^{b\bar{b}}}{N_\gamma^{t\bar{t}}}.
\label{eq:rescale}
\end{equation}
We hence use this relation to rescale Fermi-LAT expectations which are associated with the $b\bar{b}$ channel from dwarf spheroidal galaxy future data, when assuming 15 years of Fermi-LAT operation~\cite{Charles:2016pgz}.

The second conclusion from figure~\ref{fig:egamma} is that the dimension-five term in the model Lagrangian does modify the associated gamma-ray spectrum, and is therefore expected to impact on indirect detection constraints. In order to assess this further, we present in figure~\ref{fig:egamma} (right) the $\langle \sigma v \rangle$ cross section at zero velocity, indicating the contribution of the dimension-five operator to be added to the full NLO results presented in Ref.~\cite{Colucci:2018vxz}.\footnote{The study of Ref.~\cite{Colucci:2018vxz} has shown that QCD emissions play a role for high DM masses. For very small $r$, NLO corrections are hence important above approximately $m_S = 2~{\rm TeV}$, and for larger $r$ they impact the predictions for $m_S \geq 3~{\rm TeV}$.} We additionally superimpose to our results the constraints that could be extracted from Fermi-LAT and cosmic ray data and expectations. Concerning the latter, DM annihilations into $t\bar{t}$ systems can indeed be constrained with proton anti-proton cosmic ray data~\cite{Cuoco:2017iax}, given that the $t\bar{t}$ and $b\bar{b}$ spectra display the same behaviour as in figure~\ref{fig:egamma}.

The bounds achievable through the inclusion of both the dimension-five operator and NLO QCD contributions are shown in the summary plot of figure~\ref{fig:excl}, discussed in our conclusions. We recall that annihilations into pairs of gluons are neglected, as they are only relevant below threshold, for $m_S < m_t$ ({\it i.e.}\ a region into which we do not venture).

\section{Collider phenomenology} \label{sec:collider}
\begin{figure}
	\begin{center}
		\begin{tabular}{ccccc}
			\includegraphics[width=0.18\textwidth]{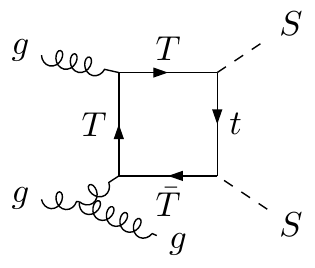}
			& \includegraphics[width=0.18\textwidth]{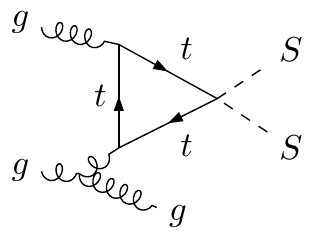}
			& \includegraphics[width=0.18\textwidth]{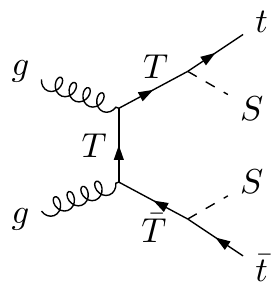}
			& \includegraphics[width=0.18\textwidth]{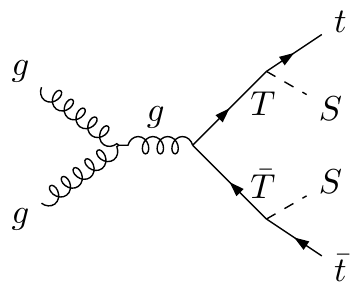}
			& \includegraphics[width=0.18\textwidth]{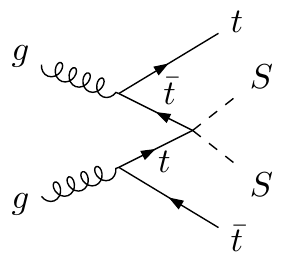}\\
			(1a) & (1b) & (2a) & (2b) & (2c)
		\end{tabular}
	\end{center}
	\caption{Processes contributing to the collider signatures of our top-philic DM model. Diagrams (1a) and (1b) yield a mono-jet signature (or a multi-jet plus missing energy one after accounting for initial-state radiation) while diagrams (2a-c)  contribute to the $t\bar{t}SS$ final state. Only  diagrams (1b) and (2c) depend on the dimension-five operator.}
	\label{fig:collamp}
\end{figure}

Finally, and in addition to modifying the astrophysical constraints, the additional vertex due to the dimension-five operator may modify the collider constraints on the model. Experimental searches for DM form an important part of the new physics search program at the LHC, and there exist a number of previous physics searches which may be reinterpreted to constrain the model examined in this article. The potential collider signatures of the model include a mono-jet channel and a multi-jet plus missing transverse energy ($\slashed{E}_T$) channel, as well as a $t\bar{t}+\slashed{E}_T$ mode. Figure~\ref{fig:collamp} shows examples of diagrams contributing to these signatures. For the mono-jet final state, diagram (1a) of figure~\ref{fig:collamp} is independent of the dimension-five operator, while diagram (1b) contains one $ttSS$ vertex. The amplitude of this latter diagram is thus proportional to $C/\Lambda$. Analogously, for the $pp\rightarrow t \bar{t} SS$ process (that leads to a multi-jet plus missing energy final state once top decays are accounted for), diagrams (2a) and (2b) of figure~\ref{fig:collamp} do not depend on the dimension-five operator, while diagram (2c) contains one $ttSS$ vertex. The full  $p p \rightarrow j S S$ and $p p \rightarrow t \bar{t} S S$  cross sections can thus be expanded as 
\begin{equation}\begin{split}
\sigma_{jSS}(m_T,m_S) =&\ \tilde{y}_t^4\hat{\sigma}^0_{jSS}(m_T,m_S)+\frac{C\tilde{y}_t^2}{\Lambda}\hat{\sigma}^{\rm{int}}_{jSS}(m_T,m_S) + \frac{C^2}{\Lambda^2}\hat{\sigma}^{\rm{dim5}}_{jSS}(m_S),\\
\sigma_{t\bar{t}SS} (m_T,m_S)=&\ \sigma^0_{t\bar{t}SS}(m_T,m_S)+\frac{C}{\Lambda}\hat{\sigma}^{\rm{int}}_{t\bar{t}SS}(m_T,m_S) + \frac{C^2}{\Lambda^2}\hat{\sigma}^{\rm{dim5}}_{t\bar{t}SS}(m_S),
\end{split}\label{eq:sigjSS}
\end{equation}
where $\sigma^0$ stand for (`bare') cross sections in the top-philic DM model \emph{without} the added dimension-five operator, $\hat{\sigma}^{\rm{int}}_{jSS}$ is the contribution from the interference of diagram (1a) with diagram (1b), $\hat{\sigma}^{\rm{int}}_{t\bar{t}SS}$ is the contribution from the interference of diagram (2a) and (2b) with diagram (2c), and $\hat{\sigma}^{\rm{dim5}}_{jSS}$ and $\hat{\sigma}^{\rm{dim5}}_{t\bar{t}SS}$ are the `bare' cross sections arising solely from the amplitude described by diagrams (1b) and (2c). For all contributions we factored out the BSM couplings $\tilde{y}_t$ and $C/\Lambda$, and indicated the dependence on the BSM particle masses $m_T$ and $m_S$.\

Obtaining the correct relic density imposes a bound of $C/\Lambda \lesssim 0.175~\rm{TeV}^{-1}$ (see our conclusions and figure~\ref{fig:comps}). The complementarity with cosmology therefore suppresses the dimension-five contributions in eq.~(\ref{eq:sigjSS}). Given $C \sim {\cal O}(1)$, we obtain effective scales of $5-100~{\rm TeV}$ for the relevant parameter space regions which yield the correct relic density. As currently pursued searches for dark matter at the LHC typically probe scales of the order of $1~{\rm TeV}$ or less, our predictions can safely be trusted in terms of the validity of the effective field theory. Moreover, $jSS$ production yields negligible cross sections once a cut as used in mono-jet searches is imposed on the jet transverse momentum ($p_T$), both for the dimension-four and dimension-five components of the cross section. We are thus left to consider the $t\bar{t}SS$ final state. For a centre-of-mass energy of $\sqrt{s}=13$~TeV, the dimension-five operator contributes to the cross section for at most 0.003~fb for $m_S > 200$~GeV, so that the inclusion of the dimension-five operator does not impact the top-philic DM model's collider phenomenology. The latter instead relies on usual vector-like top partner production, as induced by QCD interactions and whose pair-production cross section lies deep in the fb regime probed at the LHC~\cite{Fuks:2016ftf}.

To assess the LHC constraints on the model, we generate an NLO UFO model with {\sc FeynRules}~\cite{Christensen:2009jx,Alloul:2013bka,Degrande:2011ua,Degrande:2014vpa}, and then use {\sc MadGraph5\_aMC@NLO}~\cite{Alwall:2014hca} in conjunction with {\sc Pythia}~8~\cite{Sjostrand:2014zea} to generate hadron-level events for the considered process $pp\to T \bar T \to t\bar t S S$. In our simulation chain, $T$ decays are handled with {\sc MadSpin}~\cite{Artoisenet:2012st,Alwall:2014bza} and the matrix elements are convoluted with the LO set of NNPDF~3.0 set of parton densities~\cite{Ball:2014uwa,Buckley:2014ana}. The simulation of the response of the LHC detectors and event reconstruction are performed with {\sc Delphes}~3~\cite{deFavereau:2013fsa} (with appropriate detector parametrisations), that internally relies on {\sc FastJet}~\cite{Cacciari:2011ma} and its implementation of the anti-$k_T$ algorithm~\cite{Cacciari:2008gp}. In our framework both detector simulation and event reconstruction are dealt with {\sc MadAnalysis}~5~\cite{Conte:2012fm,Conte:2014zja,Dumont:2014tja,Conte:2018vmg}, which is then used for the extraction of the  CL$_s$ exclusions relative to recent ATLAS and CMS searches for dark matter in the multi-jet + $\slashed{E}_T$ and in the $t\bar t + \slashed{E}_T$ modes.

We evaluate multi-jet plus missing energy constraints by recasting the results of the ATLAS\_CONF\_2019\_040~\cite{ATLAS:2019vcq} search that covers a luminosity of $139~{\rm fb}^{-1}$ and targets final states featuring at least two hard jets in association with missing momentum. For the $t\bar t+ \slashed{E}_T$ limits, we consider the CMS-SUS-17-001 search~\cite{Sirunyan:2017leh} that analyses 35.9~fb$^{-1}$ of data and focuses on stop pair production in a final state comprising two opposite-sign isolated leptons, two hard jets, and well separated missing transverse energy. Both these analyses are available within the Public Analysis Database of {\sc MadAnalysis}~5~\cite{atlas_conf_2019,cms_sus17001}. Although the CMS-SUS-17-001 analysis has recently been updated to $137~{\rm fb}^{-1}$ of data~\cite{Sirunyan:2020tyy}, such an update is not yet included in the {\sc MadAnalysis}~5 database. However, a comparison of the observed limits in both cases reveals that there is no significant deviation when updating to the larger luminosity. We therefore follow the procedure outlined in Ref.~\cite{Araz:2019otb} to estimate the full CMS run~2 sensitivity from the CMS-SUS-17-001 one. We recalculate CL$_s$ exclusions by extrapolating the background and keeping its relative uncertainty constant, and by assuming that the new number of background events to be equal to the number of observed events.

\begin{figure}
	\includegraphics[width=0.48\textwidth]{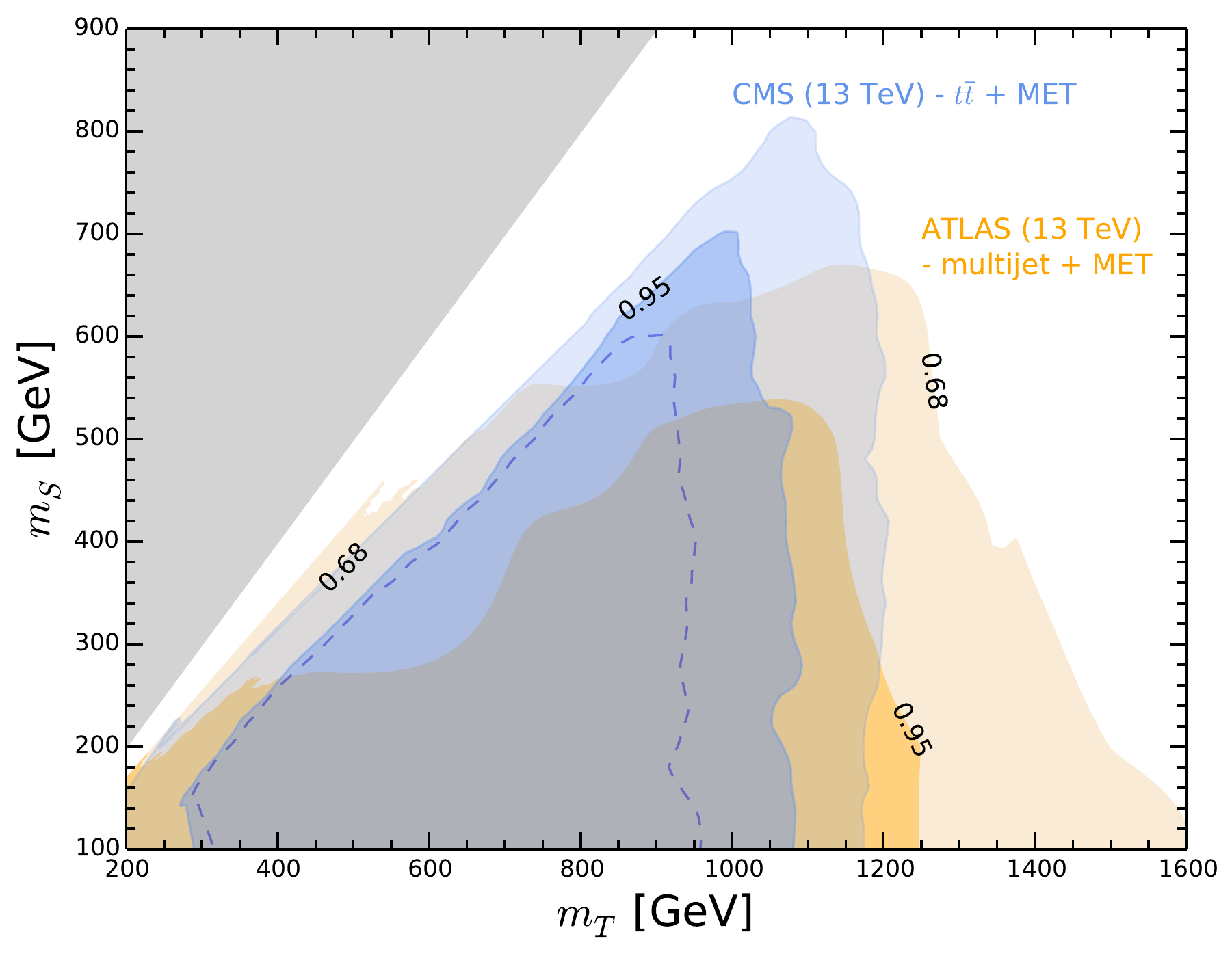}
	\hfill
	\includegraphics[width=0.48\textwidth]{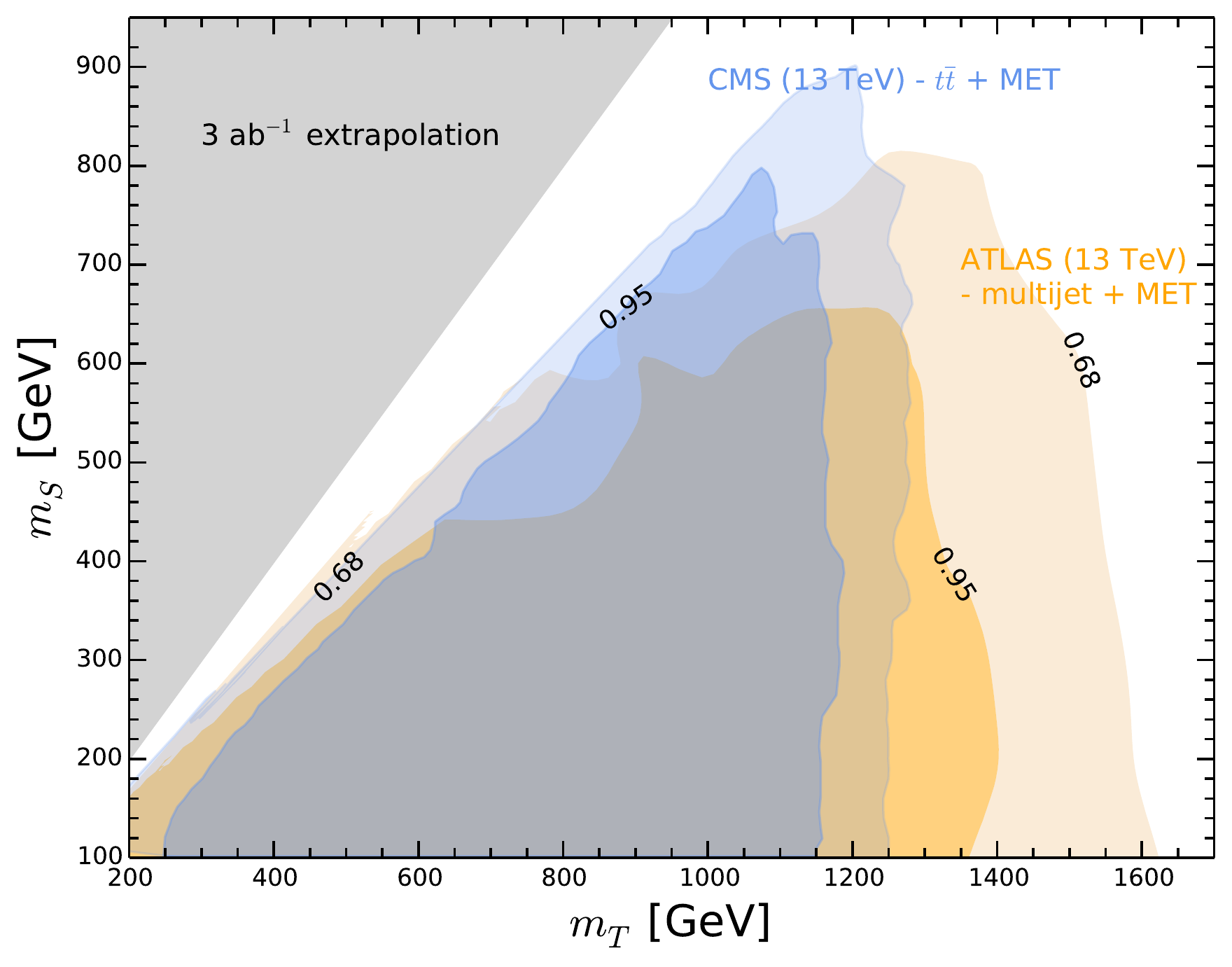}
	\caption{Collider constraints for the $t\bar{t}+\slashed{E}_T$ channel for both analyses of interest, indicating the $95~\%$ CL (darker regions) and $68~\%$ CL (lighter regions) exclusion contours. The left figure indicates LHC run-2 exclusions with a luminosity of $139~{\rm fb}^{-1}$, while the right figure presents the extrapolation of the bounds to $3~{\rm ab}^{-1}$. The dashed line featured in the left plot indicates the exclusion obtained from the CMS-SUS-17-001 analysis featuring $35.9~{\rm fb}^{-1}$ of data.}
	\label{fig:LHCexcl}
\end{figure}

The ATLAS\_CONF\_2019\_040 analysis focusing on a multi-jet final state yields relevant bounds from the $t\bar t SS$ channel. Those bounds are shown in figure~\ref{fig:LHCexcl} (left) in orange in the $(m_T, m_S)$ mass plane. In such a plane, the grey area is kinematically forbidden as the DM candidate $S$ is required to be lighter than the mediator $T$. The lighter and darker orange regions correspond to the $68\%$ and $95\%$ confidence level (CL) excluded regions respectively. As there is no information on the correlations between the different signal regions of the ATLAS analysis, we derive our exclusion bounds by solely considering the most constraining of all signal regions of the analysis. For light dark matter, mediator masses ranging up to 1.25~TeV are excluded. Such a strong bound originates from the associated split spectrum configuration that gives rise to a fair number of hard final-state jets, produced in association with a lot of missing transverse energy. Such a topology being the primary target of the ATLAS\_CONF\_2019\_040 analysis, we end up in a situation where the sensitivity is maximised. With the dark matter mass increasing, the average transverse momentum of the jet decreases once we enforce the mediator mass to be not too large so that the NLO QCD production rate stays reachable at the LHC run~2~\cite{Fuks:2016ftf}. In addition, the amount of missing transverse energy decreases accordingly, so that the sensitivity drops when the mediator mass is relatively large (too small fiducial cross section when the $p_T$ requirements on the jets are imposed) and small (too compressed spectrum) for a given $m_S$ value. For instance, for $m_S\sim 500$~GeV, we observe that mediator masses in the [800, 1100]~GeV range are excluded at 95\% CL. Furthermore, for $m_S \gtrsim 550$~GeV we lose all sensitivity. Those bounds significantly improve previous collider limits on dark matter models with a top-philic vector-like portal that are associated with the multi-jet plus $\slashed{E}_T$ search channel. The improvement corresponds to a factor of about 1.3 on the mediator mass, and to a factor of about 2 on the dark matter mass after a comparison with the bounds derived on the basis of early LHC run~2 results~\cite{Colucci:2018vxz}.

We now turn to bounds extracted from searches for the BSM production of top anti-top pairs in association with missing transverse energy. Rescaling to the full LHC run~2 luminosity the limits derived from the CMS-SUS-17-001 analysis, we present the resulting bounds as the lighter and darker blue regions of the left panel of figure~\ref{fig:LHCexcl}. These areas are excluded at 95\% and 68\% CL respectively, and we use once again the best signal region of the analysis to conservatively estimate our bounds. This analysis targeting precisely the main collider signature of the considered model ($pp\to t \bar t S S$), we can expect quite a high sensitivity to the signal. We indeed find that dark matter masses as high as about 700~GeV are reached (for a heavy mediator of about 1~TeV, so that the spectrum is not too compressed), which extends the reach of the multi-jet plus $\slashed{E}_T$ search. In addition, the $t\bar t$ plus $\slashed{E}_T$ bounds also complement the multi-jet ones in the more compressed regime, for mediator masses lying in the [400, 1000]~GeV mass window (and for dark matter being correspondingly at least 40\% lighter). For very compressed spectra, the final-state objects are not hard enough in general, so that the analysis loses sensitivity, similarly to the the multi-jet plus $\slashed{E}_T$ case. Moreover, larger mediator masses are also hard to probe due to the steeply falling NLO QCD production cross section. In order to quantify the improvements of the sensitivity relatively to the early LHC run~2 results, figure~\ref{fig:LHCexcl}  includes as a blue dashed line the early run~2 exclusion at 95\% CL that relies on a luminosity of 35.9~fb$^{-1}$. This shows that a factor of 4 in luminosity allows for stronger bounds when the mediator is heavy and the spectrum is very split, the $m_T$ lower limit increasing by about 200~GeV, for almost no change in the compressed regime.

Finally, in the right panel of figure~\ref{fig:LHCexcl}, we extrapolate our results to $3~{\rm ab}^{-1}$, which corresponds to the expected luminosity of the high-luminosity operations of the LHC. This extrapolation follows the strategy outlined above. The parameter space area that is covered extends by about 10\%--15\%, both in terms of dark matter and mediator masses when the mass spectrum is split. On the contrary, the compressed regime shows almost no improvement, as was already the case for the comparison of the bounds obtained when using a luminosity of 35.9~fb$^{-1}$ and 139~fb$^{-1}$. For these compressed scenarios, as will be discussed in the next section, cosmological probes are however in order to probe the model. A large fraction of the parameter space featuring cosmological properties in agreement with current data will however stay un-probed for the next decades. When both the dark matter and the mediator lie in the TeV or multi-TeV regime, there is indeed currently no sensitivity, either cosmologically or from colliders. Such a region being out of reach of the LHC and any planned dark matter direct or indirect detection experiment, the most fruitful strategy may be to rely on a future proton-proton collider option that would run at 100~TeV. The estimation of the corresponding reach is left for future work, where the range of validity of the effective operator should be treated with care.

\section{Summary and conclusions}\label{sec:conclusions}

\begin{figure}
	\centering
	\includegraphics[width=0.50\textwidth]{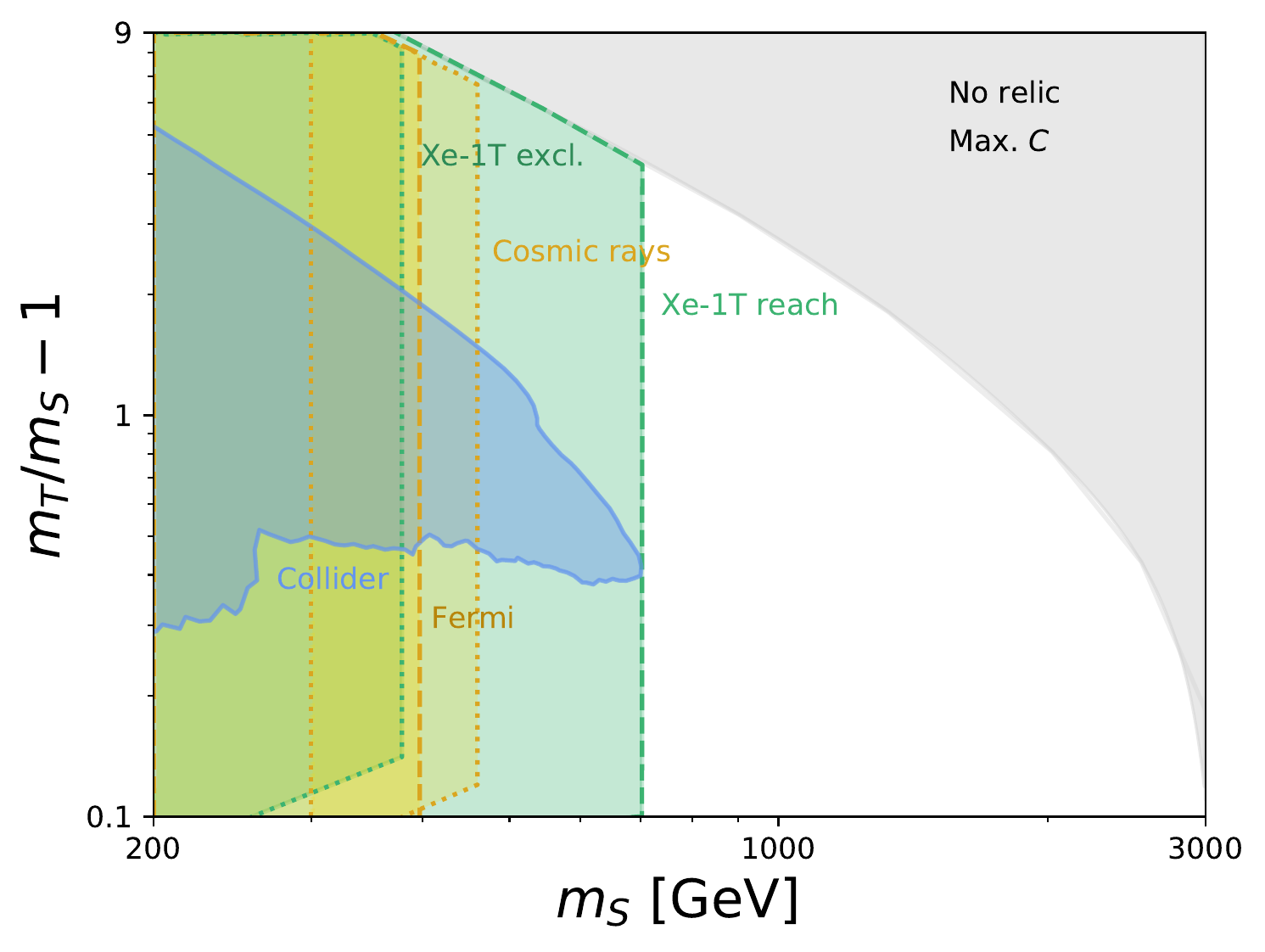}
	\includegraphics[width=0.50\textwidth]{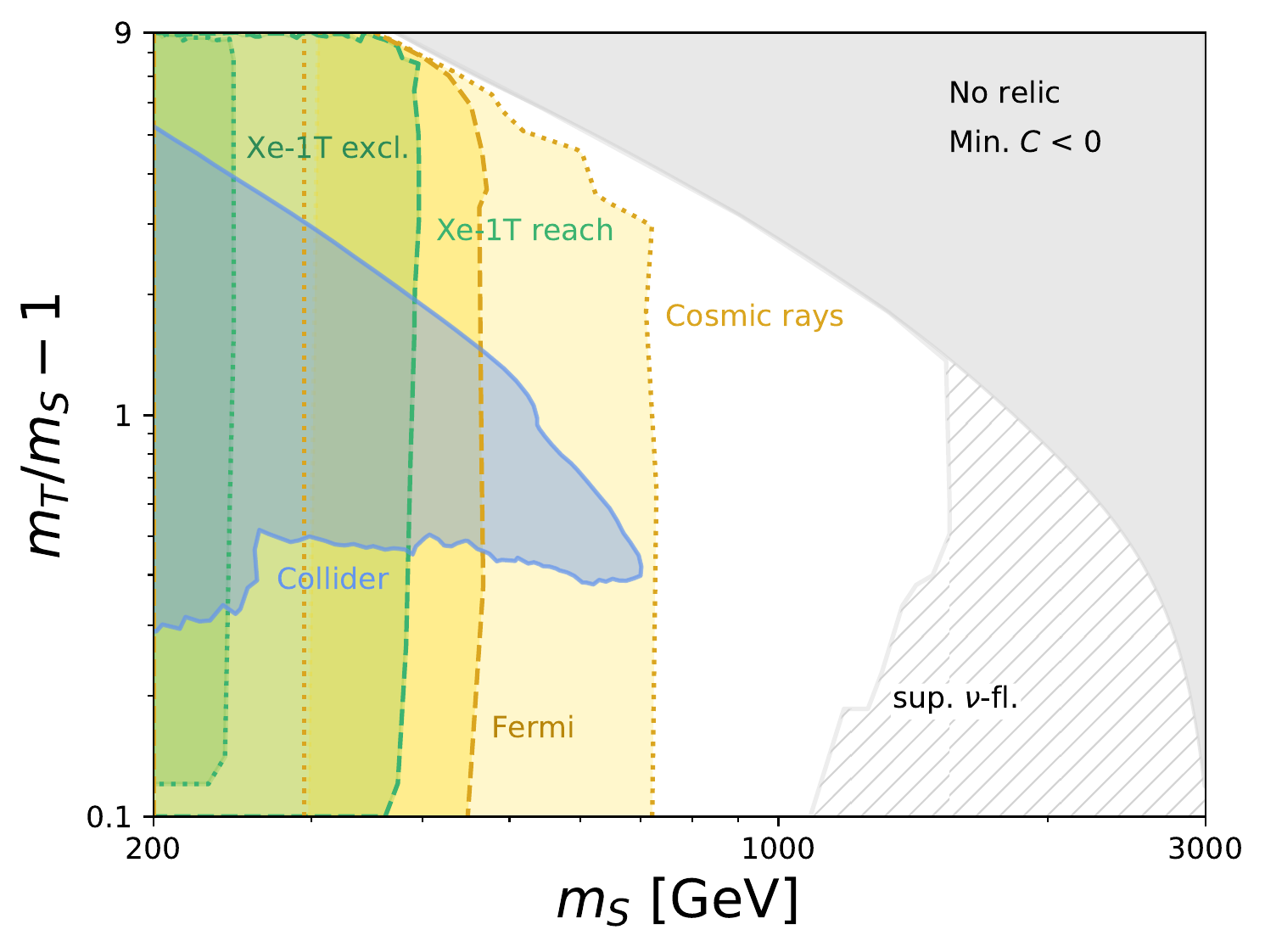}
	\includegraphics[width=0.50\textwidth]{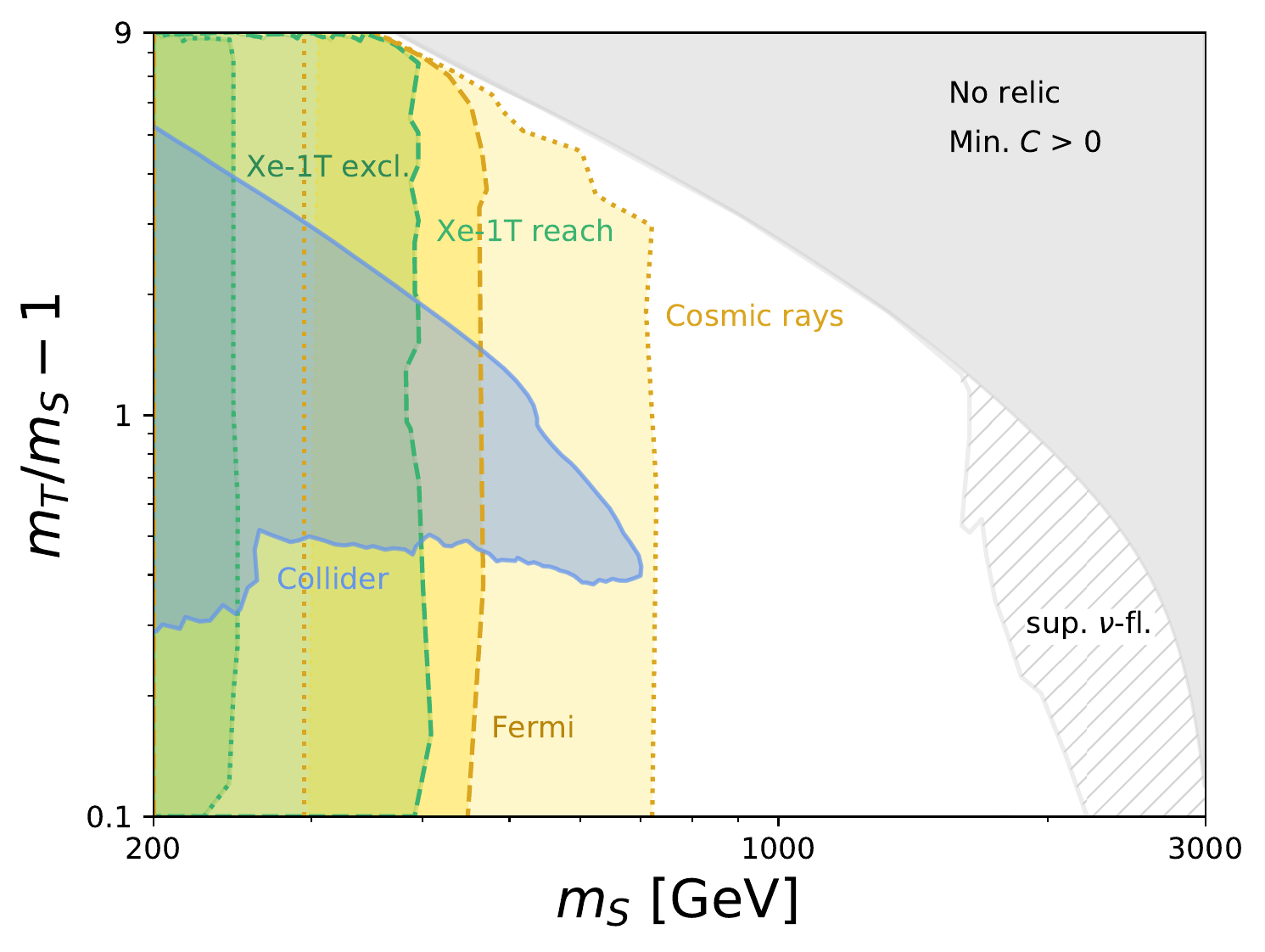}
	\caption{Exclusions and projected reaches on the considered top-philic scalar DM model. Our results are presented in the $(m_S, r=m_T/m_S-1)$ plane and include collider (blue), as well as DM direct (green) and indirect detection (yellow) constraints. For the area in grey, there is no combination of the other model parameters that yields the right relic density. For the other, viable, scenarios, we make three choices to present the results. In the top row, we consider the largest possible value of the Wilson coefficient allowing to recover the measured relic density for given $m_S$ and $m_T$ values, whilst in the central and bottom rows we choose instead the smallest $C/\Lambda$ value, either with a positive sign (lower row) or a negative one (central row). These figures summarise the findings already presented in figures~\ref{fig:DMprotonfull}, \ref{fig:egamma}, and \ref{fig:LHCexcl}.
	}
	\label{fig:excl}
\end{figure}

In this work we have examined a top-philic scalar dark matter scenario as could emerge from composite Higgs models. Our study relies on an {\it ad hoc} simplified model at the TeV scale. This model features a heavy top-philic DM candidate $S$ and a heavier vector-like fermion $T$ mediating the interactions of the dark matter with the top sector. In contrast to previous investigations, we have included not only a Yukawa coupling of the form $STt$, but also a dimension-five contact operator $ttSS$, as both are generally predicted in composite setups.

Focusing on scenarios in which the heavy vector-like top partner has a mass comparable to the DM mass, we have investigated the parameter space which yields the correct relic density numerically and semi-analytically (see section~\ref{sec:theory} and appendix~\ref{app:rdfit} for details). We have in particular examined the interplay between the contributions at dimension-four (proportional to the $STt$ Yukawa coupling) and those at dimension-five (proportional to the $ttSS$ Wilson coefficient). In investigating the direct detection constraints (see section~\ref{sec:directdetec}), we have shown that the inclusion of the dimension-five operator impacts the determination of the viable region of the scalar top-philic DM parameter space in a significant way. DM is hence allowed to be as heavy as 3 TeV in many different configurations, the corresponding direct detection cross sections being below the current XENON 1T bounds and either above (with future detection prospects) or below (and thus hard to probe) the neutrino floor. We have next studied constraints and expected bounds arising indirect DM detection in section~\ref{sec:inddec}, emphasising again the important role played by the $ttSS$ operator. Finally, current and projected LHC bounds, that are in contrast agnostic of the considered dimension-five operator, have been examined in section~\ref{sec:collider}.

We summarise our results in figure~\ref{fig:excl} which shows the various bounds and the still allowed parameter space regions. The exclusions are shown in a $(m_S, r=m_T/m_S-1)$ plane, or in other words in a plane with the dark matter mass $m_S$ and the spectrum compression factor $r$ as $x$ and $y$ axes respectively. We consider three setups. In the first one (top panel of the figure), we consider, for each pair of $S$ and $T$ masses, the largest possible value for the Wilson coefficient $C/\Lambda$ for which there exists a $\tilde y_t$ value leading to the right relic density. In the second and third considered configurations, we choose instead a scenario in which $|C/\Lambda|$ is minimum, the $\tilde y_t$ values being derived to reproduce the right relic density. We distinguish scenarios featuring a negative $C/\Lambda$ value (second panel of the figure) and a positive one (third panel of the figure). We can immediately observe the dependence of the cosmological bounds on the $C$ value, the collider bounds solely depending in contrast on the new particle masses. A larger $C$ value implies weaker cosmological bounds, whilst a minimum $C$ value reduces the impact of the direct detection bounds and increases the one of the indirect detection ones in a complementary manner. In addition, direct and indirect detection probes are the only ones relevant to enter the compressed region in which $r$ is small. Colliders have no or very little sensitivity in this regime. On the contrary, future colliders are the only way to access the so-far allowed large-mass region of the parameter space. Without such machines (whose sensitivity estimation lies beyond the scope of this work), scenarios featuring a dark matter mass larger than about 700 GeV may stay un-reachable, although they are fully viable in the light of reproducing the DM relic density as observed by Planck.

\section*{Acknowledgements}
The authors thank Laura Lopez Honorez, Michel Tytgat and J\'er\^ome Vandecasteele for their many insightful discussions and assistance, and Luca Mantani for additional discussions. ASC is supported in part by the National Research Foundation of South Africa (NRF) and thanks the University of Lyon 1 and IP2I for support during the collaboration visit in Lyon. LM is supported by the UJ GES 4IR initiative, and thanks Campus France for support under the Eiffel programme. TF’s work is supported by IBS under the project code IBS- R018-D1.

\appendix
\section{Details on the relic density fit}\label{app:rdfit}
In the following we go into further detail on the semi-analytical fit of the curve which relates the parameters $\tilde{y}_t$ and $C/\Lambda$ in producing the correct relic density. This curve depends on both $m_S$ and $m_T$ for its shape, so that imposing that the relic density matches Planck data leads to
\begin{equation}
\frac{C}{\Lambda} = f(m_S, m_T, \tilde{y}_T),
\end{equation}
where the function $f$ has to be determined.

\subsection{Without co-annihilations}
\begin{figure}
	\centering
	\includegraphics[width=0.6\textwidth]{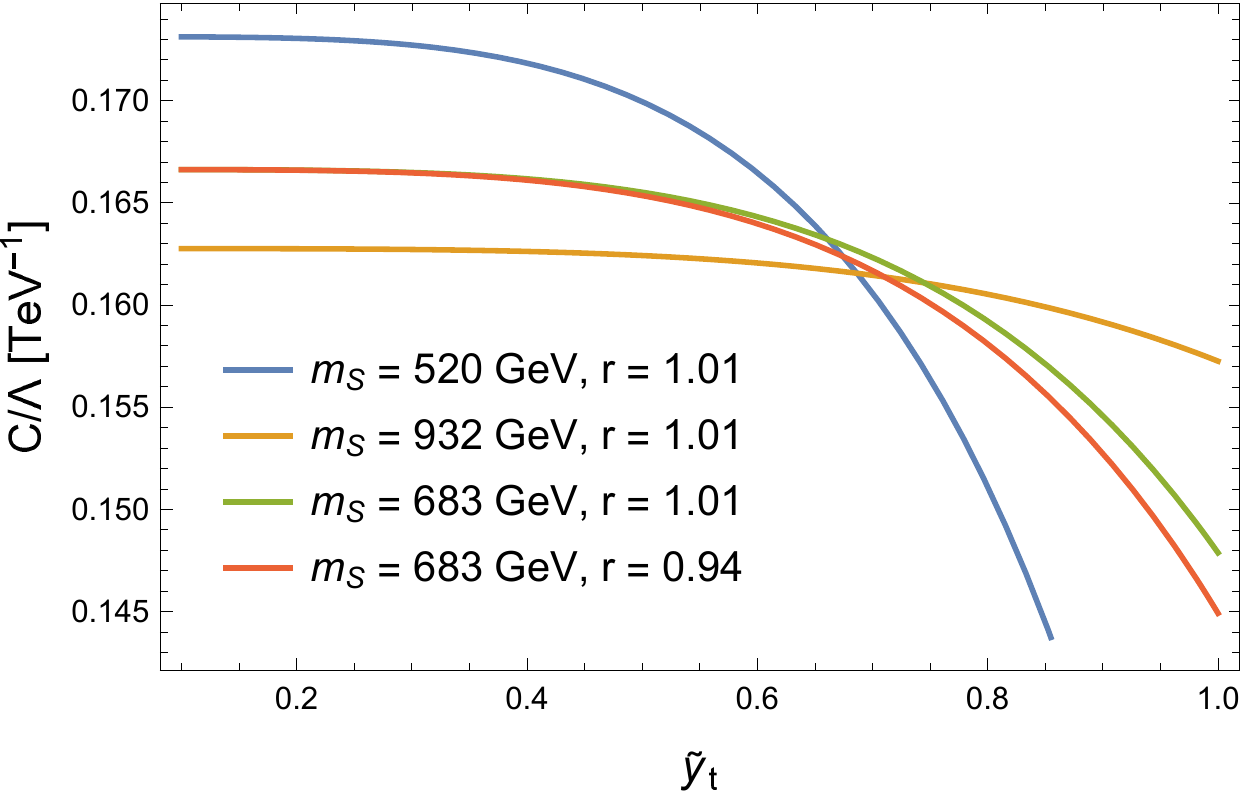}
	\caption{A comparison of the $\tilde{y}_t$ {\it versus} $C/\Lambda$ curve yielding the correct relic density for a number of mass configurations. Although a consistent shape is observable, the gradient and shift of the curve is clearly dependent on the mass parameters.}
	\label{fig:comps}
\end{figure}

We illustrate the behaviour that the $f$ function should reproduce in figure~\ref{fig:comps} for several benchmark configurations. In this figure, the behaviour of the curve is studied for approximately constant $r=m_T/m_S-1$ values across a number of $m_S$ benchmarks, and we consider scenarios for which co-annihilations are negligible. The treatment of the co-annihilation is left for the next subsection. First, we can notice that the value of $m_S$ determines the value of $C/\Lambda$ for which the dimension-five operator takes over entirely from the Yukawa coupling. In other words, for each DM mass there exists a maximum Yukawa coupling value so that $C/\Lambda$ has to be constant to reproduce the observed relic density. This is to be expected, as $\sigma v_{dim5}$ depends only on $m_S$, and not on $m_T$ as in eq.~(\ref{eq:dim5ov}). This dependence of $C/\Lambda$ on $m_S$ can be seen from the green and red lines in figure~\ref{fig:comps}, where the difference in $r$ (or equivalently on $m_T$) modifies the slopes but not the value of $C/\Lambda$ which takes over for small Yukawa values. Additionally, the figure shows that for constant $r$, a modification in $m_S$ changes both the maximal $C/\Lambda$ value and the shape of the curve.\

It is evident that the addition of the dimension-five operator $ttSS$ allows for a range of Yukawa couplings $\tilde{y}_t$ for the $STt$ operator, where previously ({\it i.e.}\ without adding the $ttSS$ operator to the Lagrangian) only a single $\tilde{y}_t$ was allowed to get the correct relic for a given mass point. In particular, we find that the addition of the dimension-five operator allows for viable scenarios featuring a relatively small Yukawa coupling. In this regime, the relic density is entirely driven by the $ttSS$ Lagrangian term. This is the first notable value of interest, where this $C/\Lambda$ lies between $0.174$ and $0.166$ TeV$^{-1}$. For smaller values of $C/\Lambda$, the contribution from the dimension-five operator to the cross section is lower, and the relic is brought back to Planck's value by the contributions involving the $STt$ operator. In short, the addition of the dimension-five operator extends the viable part of the parameter space, allowing smaller values of the Yukawa coupling $\tilde{y}_{t}$ which were previously forbidden.

For $r$ values such as the one used in figure~\ref{fig:comps}, we obtain the fitting functions shown in eq.~\eqref{eq:funcfit}.

\subsection{Parametrising the shift due to co-annihilations}

\begin{figure}
	\centering
	\includegraphics[width=0.49\textwidth]{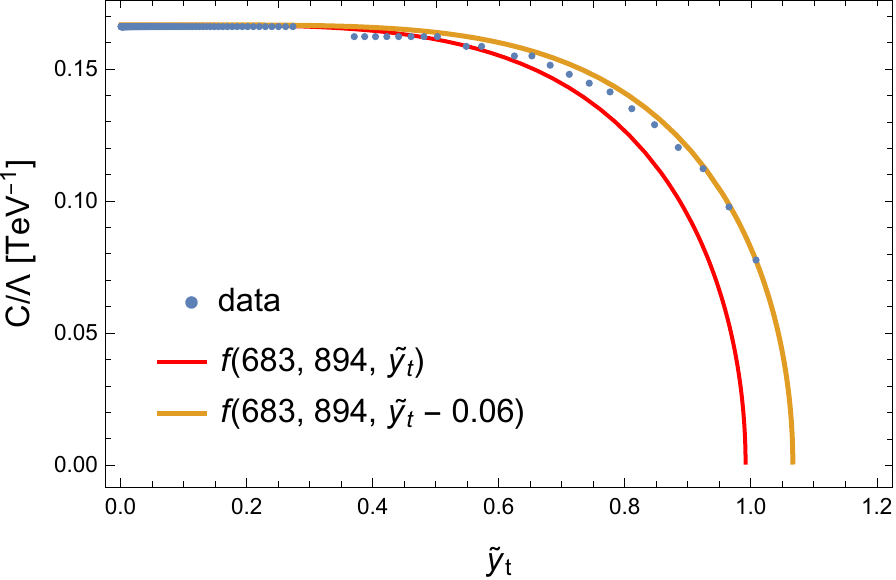}
	\hfill
	\includegraphics[width=0.49\textwidth]{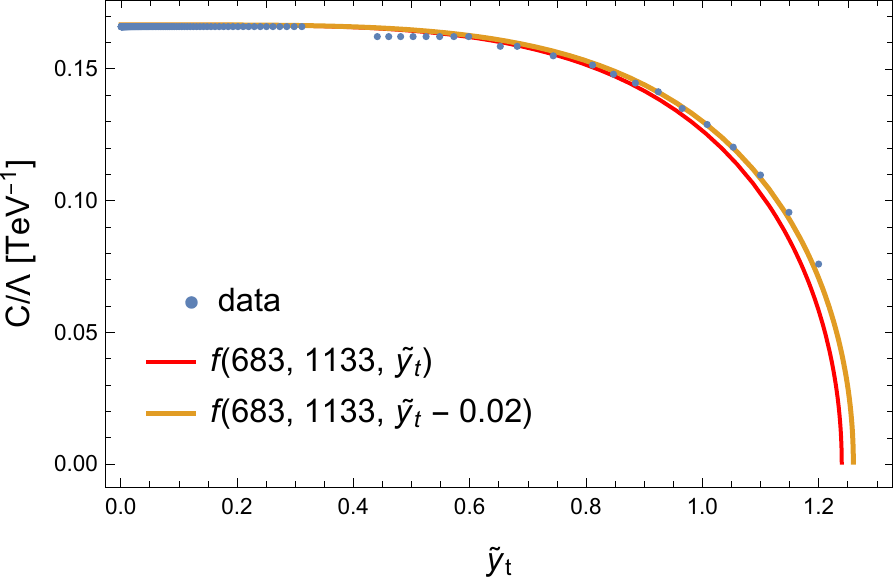}
	\hfill
	\includegraphics[width=0.49\textwidth]{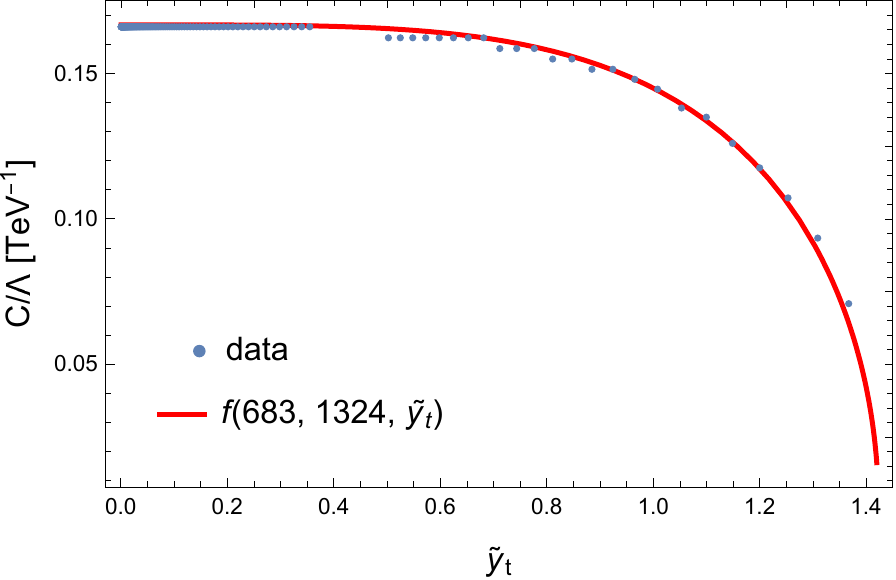}
	\hfill
	\includegraphics[width=0.46\textwidth]{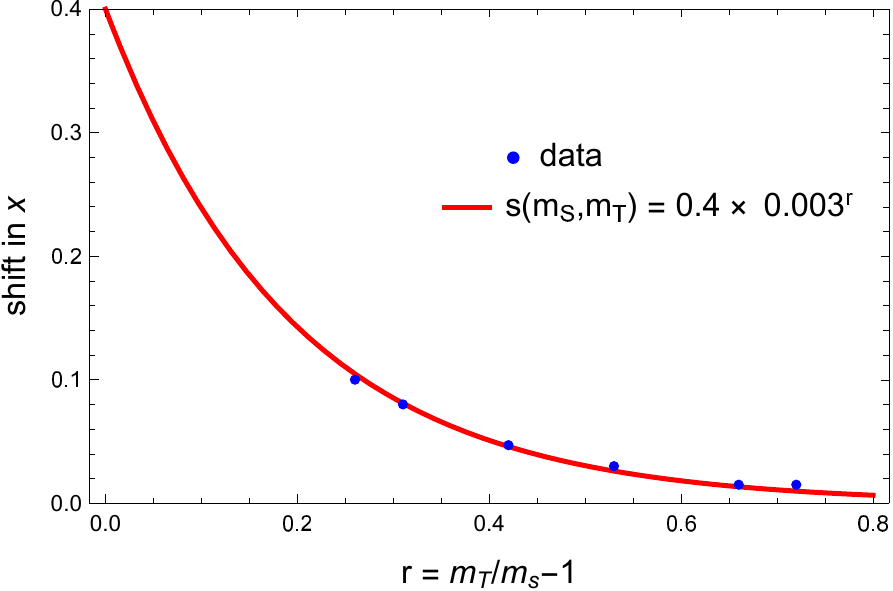}
	\caption{Results of our fitting procedure for an example mass point featuring $m_S = 682.6$~GeV, with varying $r$ values (top row and bottom left figures), and the corresponding exponential fit to the deviation (bottom right). The fitting function $f$ from eq.~\eqref{eq:funcfit} fits well for more separated masses of $S$ and $T$ (bottom left figure), but when the mass difference is small (figures of the upper row) the function deviates from the observed values. The function can be shifted in the positive $x$ direction (that is, in $\tilde{y}_t$) in order to re-establish agreement. In the bottom right plot, the deviation is fitted to a decreasing exponential, where the largest shifts in $\tilde{y}_t$ are necessary for lower $r$, and the shift becomes negligible above $r=0.8$. All points yield the correct relic density.}
	\label{fig:MS682}
\end{figure}

While for larger values of $r$ we are able to straightforwardly fit directly $C/\Lambda$ from eq.~(\ref{eq:funcfit}), for smaller values of $r$ (where $S$ and $T$ are closer in mass), we find a deviation from the fit. A small shift along the $x$-axis ({\it i.e.}\ a shift in $\tilde{y}_t$) is observed between the predictions from our scan and the fitting function. This phenomenon is visible in figure~\ref{fig:MS682}, where the red line (the initial fit) deviates from the data points ({\it i.e.}\ our numerical predictions), shown over varying $r$ for an illustrative scenario with $m_S =  682.6~{\rm GeV}$. For smaller values of $r$ (where the ``smallness" that is relevant depends on the masses of the particles), we find that the predictions are subject to a shift in $\tilde{y}_t$. We estimate this shift for the considered scenarios by the yellow lines of figure~\ref{fig:MS682}, in which the fit in shifted by some constant amount in order to agree with the relic density predictions.

The deviation between the predictions and the fitting function can be modelled simply with good agreement, as the shift follows an exponential growth as $r$ gets smaller. The amount by which the function must be shifted hence gets exponentially larger as $r$ grows smaller. This is further illustrated in the lower right panel of figure~\ref{fig:MS682}. This feature can be understood by examining the impact of co-annihilation processes on the relic density, which is relevant only when the DM candidate $S$ is close in mass with another resonance (in this case, the mediator $T$). In this scenario, the relic abundance is driven not only by self-annihilation, but also by co-annihilations between $S$ and $T$, which leads to the annihilation cross section no longer being given by the simple eq.~(\ref{eq:annihlxs}).

The calculation of the relic density must be modified to the co-annihilation case in a generalised fashion~\cite{Griest:1990kh, Kolb:1990vq, Servant:2002aq}. The Boltzmann equation~(\ref{eq:boltzmann}) is generalised to a set of coupled equations governing the evolution of the $S$ and $T$ states through the universe's history. Focusing on the dark matter (co-)annihilation cross section only, we have
\begin{equation}
\sigma_{eff}(x) =  \sigma_{SS} +  \sigma_{ST}\ \frac{g_Sg_T}{g_{eff}^2}\ 
\bigg(\frac{m_T}{m_S}\bigg)^{3/2}\ {\rm exp}\big[-x~r\big],
\end{equation}
where $x = m_S/T^*$ with $T^*$ being the temperature, and where $g_S$ and $g_T$ stand for the DM and mediator number of internal degrees of freedom, and $g_{eff}$ for the effective number of internal degrees of freedom. Moreover, $\sigma_{SS}$ and $\sigma_{ST}$ respectively correspond to the annihilation ($SS\to t\bar t$) and co-annihilation ($ST\to t^{(*)}\to X$) cross sections. It is then apparent that we can expect the deviation from the fit to be exponentially larger for smaller $r$, and we can approximate this deviation using an exponential function.\

\begin{figure}
	\centering
	\includegraphics[width=0.49\textwidth]{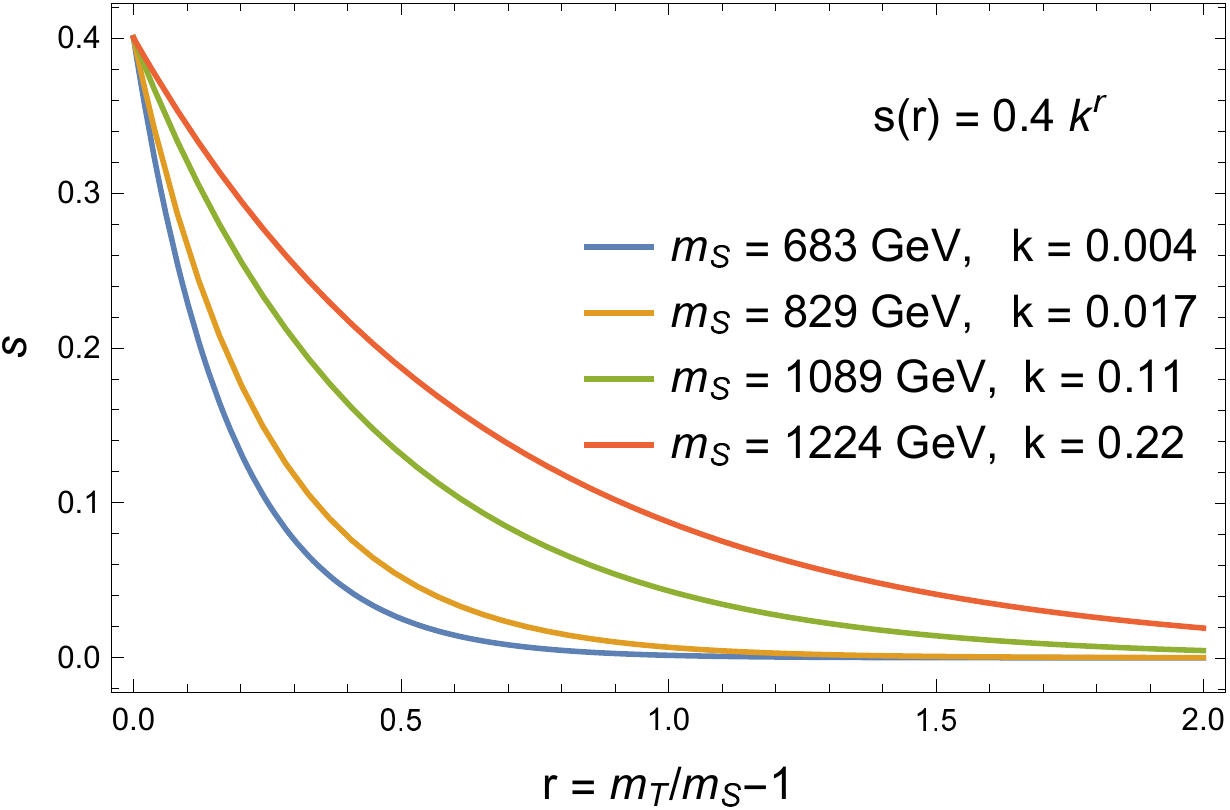}
	\hfill
	\includegraphics[width=0.49\textwidth]{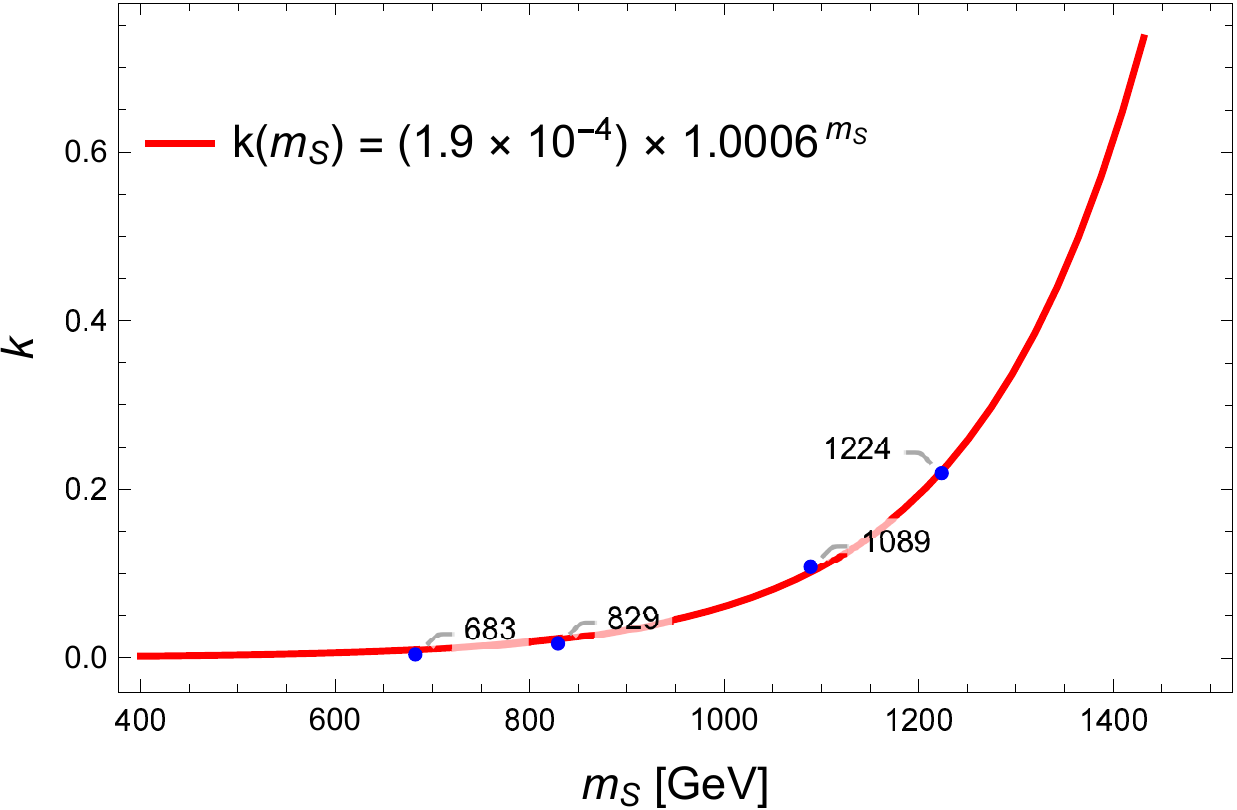}
	\hfill
	\includegraphics[width=0.49\textwidth]{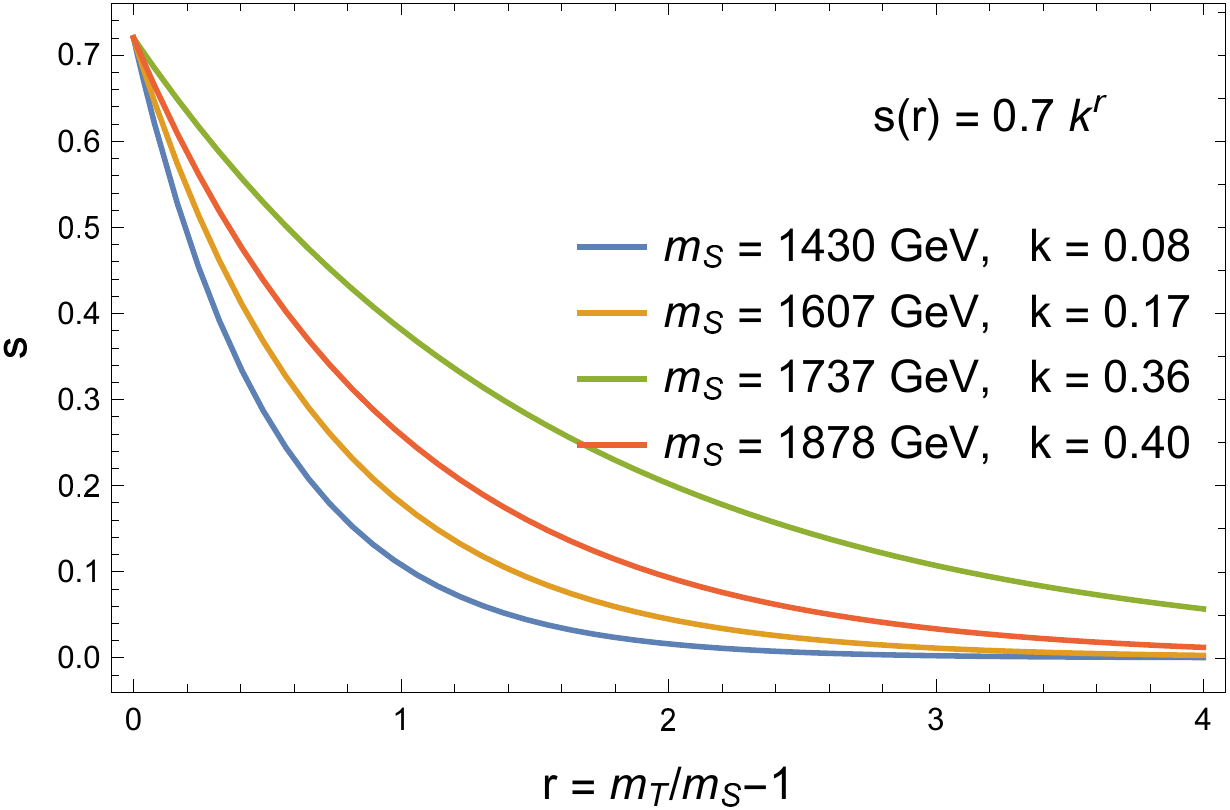}
	\hfill
	\includegraphics[width=0.49\textwidth]{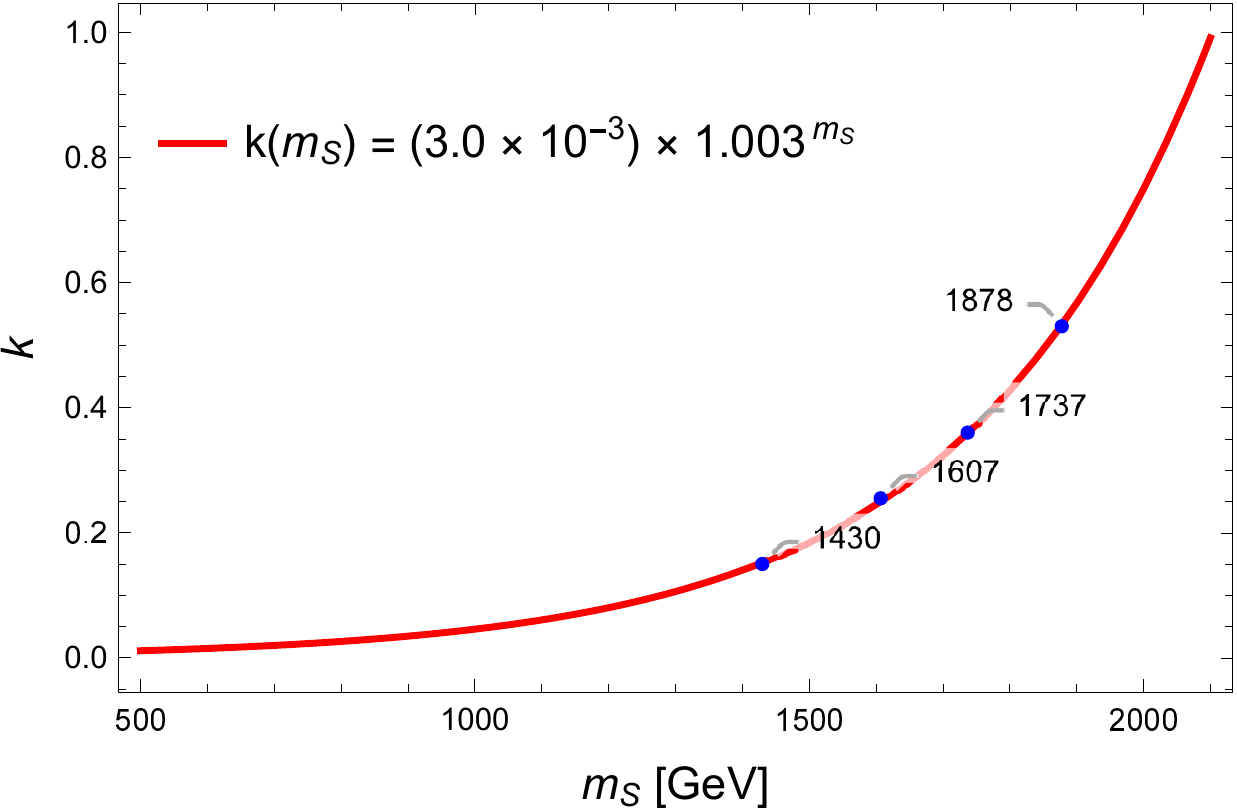}
	\caption{{\it Left figures} -- Deviation function~\eqref{eq:shiftdeviation} for several $m_S$ benchmark points. it parametrises the deviations between the fit function~\eqref{eq:funcfit} and our predictions in the compressed regime, and it is shown as a function of $r$. {\it Right figures} -- Value of the $k$ parameter driving the deviations from the fit across the considered mass points. This exhibits again an exponential structure. }
	\label{fig:allshifts}
\end{figure}

Across the range of allowed masses for the DM candidate, two separate regimes are observed in the behaviour of the interplay of the different contributions to the annihilation cross section. Below $m_S \approx 1~{\rm TeV}$, the NLO effects are not yet dominant, and $\sigma_{NLO}/ \sigma_{vqq} \approx 1$. For higher mass regions, the NLO cross section is influenced by the VIB contributions, as apparent in the example benchmarks in figure~\ref{fig:comparexsecs}. This motivates distinguishing two mass regimes, above and below about $1~{\rm TeV}$. At each $m_S$ benchmark we hence find a smooth pattern of shifts across $r$ which can be fitted to an exponential function. Repeating the fit done in figure~\ref{fig:MS682} across a number of benchmarks, the function which determines the shift $s$ for a mass point $(m_S, m_T)$ is found to take the form
\begin{equation}
s(m_S,m_T) = \begin{cases} 0.4 k^r & m_S\leq 1.2~{\rm TeV}\\
0.7 k^r & m_S > 1.2~{\rm TeV} \\
\end{cases}\ ,\label{eq:shiftdeviation}
\end{equation}
where $k$ is an unconstrained parameter of dimension 1 allowed to vary across the benchmarks (but constant for a given $m_S$). The constants are determined by the fit. In figure~\ref{fig:allshifts} (left), we show these exponential functions which map the shifts for a number of $m_S$ benchmarks. \

Finally, we would like to be able to estimate the value of the parameter $k$ in the exponential shift function, as it is specific to the mass $m_S$. We find that the values of this also may be fitted to an exponential function fully determined by $m_S$, as displayed in figure~\ref{fig:allshifts}. We obtain
\begin{equation}
k(m_S) =  \begin{cases} (1.9\times 10^{-4}) \left(6.2\times 10^8\right)^{m_S/ \Lambda} & m_S\leq 1.2~{\rm TeV}\\
(3.0\times 10^{-3}) (1.8\times 10^4)^{m_S/ \Lambda} & m_S > 1.2~{\rm TeV} \\
\end{cases}\ ,
\end{equation}
for $\Lambda = 3.5~{\rm TeV}$. The value of the dimension-five coupling contributing to the relic density for a given benchmark $(m_S, m_T)$ can then be fully determined by extending eq.~(\ref{eq:funcfit}) to include the shift, as previously quoted in eq.~(\ref{eq:full}).


\bibliographystyle{JHEP-2-2}
\bibliography{cdmbib}

\end{document}